\documentclass[twocolumn,epjc3]{svjour3mod}  
\journalname{Eur. Phys. J. C}
\usepackage[utf8]{inputenc}
\usepackage{epsfig,amsfonts,amssymb,latexsym,amsmath,bm,color,array,epsfig,graphics,dsfont}
\usepackage{enumitem}
\usepackage{amsmath}
\usepackage{amssymb}
\usepackage{physics} 
\usepackage{lastpage}
\usepackage{graphicx}
\usepackage[colorlinks,citecolor=blue,urlcolor=blue,linkcolor=blue,breaklinks]{hyperref}
\usepackage{balance,cite}
\usepackage{booktabs}
\usepackage{multirow}
\usepackage{subcaption}
\newcommand{\mpi}{M_\pi}

\newcommand{\metap}{M_{\eta^\prime}}
\newcommand{\Getap}{\Gamma_{\eta^\prime}}
\newcommand{\beq}{\begin{equation}}
\newcommand{\eeq}{\end{equation}}
\newcommand{\diff}{\text{d}}
\newcommand{\eps}{\epsilon}
\newcommand{\Br}{\text{Br}}

\newcommand{\Fpi}{F_\pi}

\newcommand{\Order}{\mathcal{O}}
\newcommand{\M}{\mathcal{M}}

\newcommand{\me}{m_e}
\newcommand{\mw}{M_\omega}
\newcommand{\mwbare}{M_{\omega,0}}
\newcommand{\Gw}{\Gamma_\omega}
\newcommand{\mwb}{\bar M_\omega}
\newcommand{\Gwb}{\bar\Gamma_\omega}
\newcommand{\gw}{g_{\omega 3}}
\newcommand{\gww}{g_{\omega 2}}
\newcommand{\gwg}{g_{\omega\gamma}}
\newcommand{\gwb}{\bar g_{\omega 3}}

\newcommand{\gwgb}{\bar g_{\omega\gamma}}
\newcommand{\epsrw}{\eps_{\rho\omega}}

\newcommand{\Lagr}{\mathcal{L}}
\newcommand{\unity}{\mathds{1}}
\newcommand{\F}{\mathcal{F}}
\newcommand{\mr}{M_\rho}
\newcommand{\Gr}{\Gamma_\rho}
\newcommand{\gr}{g_{\rho\pi\pi}}
\newcommand{\grg}{g_{\rho\gamma}}
\newcommand{\Sw}{\Sigma_{3\pi}}
\newcommand{\MP}{M_P}
\newcommand{\mphi}{M_\phi}
\newcommand{\Gphi}{\Gamma_\phi}
\newcommand{\GeV}{\,\text{GeV}}
\newcommand{\MeV}{\,\text{MeV}}
\newcommand{\keV}{\,\text{keV}}
\renewcommand{\Im}{\text{Im}\,}

\newcommand{\Impipi}{\text{Im}_{\pi\pi}\,}
\newcommand{\Impipipi}{\text{Im}_{3\pi}\,}

\allowdisplaybreaks[1]

\numberwithin{equation}{section}

\begin{document}
\emergencystretch 3em
\title{A dispersive analysis of $\boldsymbol{\eta'\to\pi^+\pi^-\gamma}$ and $\boldsymbol{\eta'\to \ell^+\ell^-\gamma}$}
\author{ 
Simon Holz\thanksref{add1,add3,e1} 
\and Christoph Hanhart\thanksref{add2}
\and Martin Hoferichter\thanksref{add3,add4}
\and 
Bastian~Kubis\thanksref{add1}
}       
\thankstext{e1}{e-mail: holz@itp.unibe.ch}
%

\institute
{Helmholtz-Institut f\"ur Strahlen- und Kernphysik and Bethe Center for Theoretical Physics, Universit\"at Bonn, 53115 Bonn, Germany \label{add1}
\and
Albert Einstein Center for Fundamental Physics, Institute for Theoretical Physics, University of Bern, Sidlerstrasse 5, 3012 Bern, Switzerland \label{add3}
\and
Forschungszentrum J\"ulich, Institute for Advanced Simulation, Institut f\"ur Kernphysik, and
J\"ulich Center for Hadron Physics, 52425 J\"ulich, Germany \label{add2}
\and 
Institute for Nuclear Theory, University of Washington, Seattle, WA 98195-1550, USA \label{add4}
}
\date{}
\maketitle
\begin{abstract}
We present a dispersive representation of the $\eta'$ transition form factor that allows one to account, in a consistent way, for the effects of $\rho$--$\omega$ mixing in both the isoscalar and the isovector contributions. Using this formalism, we analyze recent data on $\eta'\to \pi^+\pi^-\gamma$ to constrain the isovector part of the form factor, individually and in combination with data for the pion vector form factor.  
As a first application, we use our results, in combination with the most recent input for the isoscalar part of the form factor, to 
predict the corresponding spectrum of $\eta'\to\ell^+\ell^-\gamma$, in particular we find the slope parameter $b_{\eta'}=1.431(23)\GeV^{-2}$. 
With forthcoming data on the latter process, our results establish the necessary framework to improve the evaluation of the $\eta'$-pole contribution to the anomalous magnetic moment of the muon using experimental input from both $\eta'$ decay channels.  
\end{abstract}
\section{Introduction}

The transition form factors (TFFs) of pseudoscalar mesons, $F_{P\gamma^*\gamma^*}(q_1^2,q_2^2)$ with $P=\pi^0,\eta,\eta'$, describe the interaction with two (virtual) photons,
\begin{align}
\label{TFF_def}
&i\int\diff^4 x\,e^{i q_1\cdot x}\big\langle 0\big|T\big\{j_\mu(x)j_\nu(0)\big\}\big|P(q_1+q_2)\big\rangle\notag\\
& \hspace{3cm}=\eps_{\mu\nu\alpha\beta}q_1^\alpha q_2^\beta F_{P\gamma^*\gamma^*}(q_1^2,q_2^2),
\end{align}
where $j_\mu = (2\bar{u}\gamma_\mu u-\bar{d}\gamma_\mu d-\bar{s}\gamma_\mu s)/3$ is the electromagnetic current, $q_{1,2}$ are the photon momenta, and $\eps^{0123}=+1$. For both photons on-shell, these form factors determine the di-photon decays governed by the chiral anomaly~\cite{Wess:1971yu,Witten:1983tw}
\beq
\Gamma(P\to\gamma\gamma)=\frac{\pi\alpha^2 \MP^3}{4}\big|F_{P\gamma\gamma}\big|^2,
\eeq
with $F_{P\gamma\gamma}=F_{P\gamma^*\gamma^*}(0,0)$. For the pion, the corresponding normalization was experimentally~\cite{PrimEx-II:2020jwd} found to be close to the prediction from the low-energy theorem~\cite{Adler:1969gk,Bell:1969ts,Bardeen:1969md} $F_{\pi^0\gamma\gamma}=1/(4\pi^2\Fpi)$ in terms of the pion decay constant $\Fpi$, to the extent that higher-order corrections~\cite{Bijnens:1988kx,Goity:2002nn,Ananthanarayan:2002kj,Kampf:2009tk} thwart agreement with experiment. For $P=\eta,\eta'$ the analog relations depend on the details of $\eta$--$\eta'$ mixing~\cite{Leutwyler:1997yr,Feldmann:1998vh,Feldmann:1998sh,Feldmann:1999uf,Escribano:2005qq,Escribano:2015yup,Ottnad:2017bjt}, see Ref.~\cite{Gan:2020aco} for a review. Here, we will use the outcome of the PDG global fit, $\Gamma(\eta'\to\gamma\gamma)=4.34(14)\keV$~\cite{ParticleDataGroup:2020ssz}, i.e.,
\beq
F_{\eta'\gamma\gamma}=0.3437(55)\GeV^{-1}.
\eeq

Beyond the normalization, understanding the momentum dependence of the TFFs is critical to be able to calculate the pseudoscalar-pole contributions to hadronic light-by-light scattering (HLbL) in the anomalous magnetic moment of the muon. Currently, the main uncertainty in the Standard-Model prediction~\cite{Aoyama:2020ynm,Aoyama:2012wk,Aoyama:2019ryr,Czarnecki:2002nt,Gnendiger:2013pva,Davier:2017zfy,Keshavarzi:2018mgv,Colangelo:2018mtw,Hoferichter:2019mqg,Davier:2019can,Keshavarzi:2019abf,Hoid:2020xjs,Kurz:2014wya,Melnikov:2003xd,Colangelo:2014dfa,Colangelo:2014pva,Colangelo:2015ama,Masjuan:2017tvw,Colangelo:2017qdm,Colangelo:2017fiz,Hoferichter:2018dmo,Hoferichter:2018kwz,Gerardin:2019vio,Bijnens:2019ghy,Colangelo:2019lpu,Colangelo:2019uex,Blum:2019ugy,Colangelo:2014qya,Hoferichter:2021wyj}
\beq
\label{amuSM}
a_\mu^\text{SM}=116\,591\,810(43)\times 10^{-11} 
\eeq
originates from hadronic vacuum polarization, see, e.g., Refs.~\cite{Aoyama:2020ynm,Borsanyi:2020mff,Lehner:2020crt,Crivellin:2020zul,Keshavarzi:2020bfy,Malaescu:2020zuc,Colangelo:2020lcg} for further discussion, but to match the final projected precision of the Fermilab experiment, which will improve upon the current world average 
~\cite{bennett:2006fi,Abi:2021gix,Albahri:2021ixb,Albahri:2021kmg,Albahri:2021mtf}
\beq
\label{exp}
a_\mu^\text{exp}=116\,592\,061(41)\times 10^{-11}
\eeq
by more than another factor of two, also the uncertainties in the subleading HLbL contribution, in Ref.~\cite{Aoyama:2020ynm} estimated as
~\cite{Melnikov:2003xd,Colangelo:2014dfa,Colangelo:2014pva,Colangelo:2015ama,Masjuan:2017tvw,Colangelo:2017qdm,Colangelo:2017fiz,Hoferichter:2018dmo,Hoferichter:2018kwz,Gerardin:2019vio,Bijnens:2019ghy,Colangelo:2019lpu,Colangelo:2019uex,Pauk:2014rta,Danilkin:2016hnh,Jegerlehner:2017gek,Knecht:2018sci,Eichmann:2019bqf,Roig:2019reh,Blum:2019ugy}
\begin{equation}
\label{HLbL}
 a_\mu^\text{HLbL}=90(17)\times
10^{-11},
\end{equation}
need to be reduced accordingly. Recent progress includes a second complete lattice calculation~\cite{Chao:2021tvp}, while on the phenomenological side the role of higher intermediate states and the implementation of short-distance constraints are being scrutinized~\cite{Leutgeb:2019gbz,Cappiello:2019hwh,Hoferichter:2020lap,Ludtke:2020moa,Bijnens:2020xnl,Bijnens:2021jqo,Zanke:2021wiq,Danilkin:2021icn,Colangelo:2021nkr}. 
Besides these subleading contributions, the $\eta$ and $\eta'$ poles are currently estimated using Canterbury approximants alone~\cite{Masjuan:2017tvw}---as opposed to the $\pi^0$ pole, for which independent calculations from dispersion relations~\cite{Hoferichter:2018dmo,Hoferichter:2018kwz}, lattice QCD~\cite{Gerardin:2019vio}, and Canterbury approximants all give a coherent picture---so that a full dispersive analysis is called for to corroborate the corresponding uncertainty estimates.  Several steps in this direction have already been taken in previous work~\cite{Stollenwerk:2011zz,Hanhart:2013vba,Kubis:2015sga,Holz:2015tcg}, in particular, towards a better understanding of the role of factorization-breaking terms~\cite{Holz:2015tcg}. 

In this paper, we address another subtlety that is related to the interplay of isoscalar and isovector contributions. In principle, since $\eta$, $\eta'$ have isospin $I=0$, both photons need to be either isoscalar or isovector, leading to a simple vector-meson-dominance (VMD) picture of decays proceedings either via two $\rho$ mesons or some combination of $\omega$ and $\phi$. However, isospin-breaking effects are resonance enhanced just as in the pion vector form factor (VFF), so that both the admixture of an $\omega$ into the isovector contribution and, vice versa, of $\pi\pi$ intermediate states into the isoscalar component can become phenomenologically relevant, at least in the vicinity of the resonance. This issue becomes particularly important when the two-pion cut in the isovector contribution is constrained via data for $\eta'\to\pi^+\pi^-\gamma$, since also in this case isoscalar corrections will enter. Based on Ref.~\cite{Hanhart:2012wi} we develop a coupled-channel formalism that allows one to disentangle these effects in a consistent manner, apply the result to recent $\eta'\to\pi^+\pi^-\gamma$ data from BESIII~\cite{BESIII:2017kyd}, and then calculate the resulting singly-virtual TFF to predict the spectrum for $\eta'\to\ell^+\ell^-\gamma$. A summary of the formalism is given in Sec.~\ref{sec:formalism}, with a detailed derivation in the appendix. Fits to the $\eta'\to\pi^+\pi^-\gamma$ data are presented in Sec.~\ref{sec:etap_ppg}, followed by the prediction for $\eta'\to\ell^+\ell^-\gamma$ in Sec.~\ref{sec:etap_llg} and our conclusions in Sec.~\ref{sec:conclusions}.

\section{Formalism}
\label{sec:formalism}

The singly-virtual TFF in the definition~\eqref{TFF_def} determines the spectrum for $P\to\ell^+\ell^-\gamma$ according to
\begin{align}
\label{decay_eeg}
\frac{\diff\Gamma(P\to \ell^+\ell^-\gamma)}{\diff s}&=\frac{\alpha^3(\MP^2-s)^3(s+2m_\ell^2)\sigma_\ell(s)}{6\MP^3s^2|1-\Pi(s)|^2}\notag\\
&\quad\times\big|F_{P\gamma^*\gamma^*}(s,0)\big|^2,
\end{align}
where $s$ is the invariant mass of the lepton pair and $\sigma_\ell(s)=\sqrt{1-4m_\ell^2/s}$ the phase-space variable.\footnote{We define the form factor $F_{P\gamma^*\gamma^*}$ excluding vacuum-polarization (VP) corrections. Unless removed in the experimental analysis---in analogy, e.g., to the bare cross section for $e^+e^-\to\pi^+\pi^-$---this correction $1/(1-\Pi(s))$ therefore appears in the expression~\eqref{decay_eeg} for the decay width.} Once $F_{P\gamma^*\gamma^*}(s,0)$ is known, the di-lepton spectrum can thus be predicted. 

However, the differential decay width~\eqref{decay_eeg} scales as $\Order(\alpha^3)$ in the fine-structure constant $\alpha=e^2/(4\pi)$, leading to a challenging experimental signature. Additional information on the energy dependence can be obtained by combining the $P\to \pi^+\pi^-\gamma$ decay with the pion VFF $F_\pi^V$,
\beq
\label{eq:pvff_def}
	\langle \pi^\pm(p') | j^\mu(0) | \pi^\pm(p) \rangle =\pm (p'+p)^\mu F_\pi^V((p'-p)^2),
\eeq
which determines the discontinuity of the dominant $2\pi$ intermediate states. Experimentally, the decay 
$P\to\pi^+\pi^-\gamma$ is more easily accessible, given that in this case the differential decay width only scales as $\Order(\alpha)$, and information on the spectrum can then be used to reconstruct the isovector part of the TFF. This strategy is straightforward as long as isospin violation is neglected, while the transition from $P\to\pi^+\pi^-\gamma$ to $P\to\ell^+\ell^-\gamma$ becomes more intricate once such corrections are included. In particular, $\rho$--$\omega$ mixing is enhanced by the presence of the resonance propagator, and therefore needs to be included to obtain a realistic line shape. In such a situation, one cannot consider the $2\pi$ channel in isolation anymore, since also $F_\pi^V$ depends on $\rho$--$\omega$ mixing, leading to a spectral function in which the double discontinuities of $2\pi$ and $3\pi$ intermediate states, corresponding to $\rho$ and $\omega$, respectively, no longer cancel. In~\ref{app:coupled}--\ref{app:full} we systematically develop a formalism that avoids such inconsistencies, allowing for a meaningful consideration of $\rho$--$\omega$ mixing in both the isoscalar and isovector contributions. The central result is given by
\begin{align}
\label{rep_final}
 F_{P\gamma^*\gamma^*}(s,0)&=F_{P\gamma\gamma}+
 \bigg[1+\frac{\epsrw s}{\mw^2-s-i\mw\Gw}\bigg]\notag\\
 &\quad\times \frac{s}{48\pi^2}\int_{4\mpi^2}^\infty\diff s'\frac{\sigma_\pi^3(s')P(s')|F_\pi^V(s')|^2}{s'-s-i\eps}\notag\\
 &+\frac{F_{P\gamma\gamma}w_{P\omega\gamma}s}{\mw^2-s-i\mw\Gw}\bigg[1+\frac{\epsrw s}{48\pi^2\gwg^2}
 \notag\\
 &\quad\times\int_{4\mpi^2}^\infty\diff s'\frac{\sigma_\pi^3(s')|F_\pi^V(s')|^2}{s'(s'-s-i\eps)}\bigg] \notag\\
 &+\frac{F_{P\gamma\gamma}w_{P\phi\gamma}s}{\mphi^2-s-i\mphi\Gphi},
\end{align}
a dispersion relation constructed from a spectral function whose double discontinuity vanishes, see Eq.~\eqref{double_cancel}. Equation~\eqref{rep_final} is expressed in terms of the $\rho$--$\omega$ mixing parameter $\epsrw$ and the weights $w_{PV\gamma}$, defined in Eq.~\eqref{weights}, which determine the isoscalar contribution to the slope of the TFF, as well as the second-order polynomial 
\begin{equation}
 P(s) = \frac{A}{2}\big(1+\beta s+\gamma s^2\big)   \label{defP}
\end{equation}
introduced to describe the $\eta' \to \pi^+ \pi^- \gamma$ decay spectrum below.

\section{Fits to $\eta'\to\pi^+\pi^-\gamma$ and $e^+ e^- \to \pi^+ \pi^-$}
\label{sec:etap_ppg}

For the pion VFF, defined in Eq.~\eqref{eq:pvff_def}, we employ a dispersive representation
\beq
\label{eq:pvff_omnes}
    F_\pi^V(s) = (1 + \alpha_\pi s) \Omega(s),
\eeq
where
\beq
    \Omega(s) = \exp \left\lbrace \frac{s}{\pi} \int_{4\mpi^2}^{\infty} \diff x\, \frac{\delta_1^1(x)}{x(x-s-i \eps)} \right\rbrace
\eeq
is the Omn\`es function~\cite{Omnes:1958hv} and $\delta_1^1(s)$ denotes the $\pi \pi$ $P$-wave scattering phase shift in the isospin $I=1$ channel. As input for the phase shift we use the solution of the Roy-equation analysis optimized for fits to pion VFF data of Ref.~\cite{Colangelo:2018mtw}. The term multiplying the Omn\`es function in Eq.~\eqref{eq:pvff_omnes} takes the effects of inelastic contributions, such as $4 \pi$, as well as our ignorance about the high-energy behavior of the phase shift into account, where the constant $\alpha_\pi$ is left as a free parameter to be constrained by the fit. The isospin-breaking effect of $\rho$--$\omega$ mixing in $e^+ e^- \to \pi^+ \pi^-$ is parameterized via
\beq
    F_\pi^{V,e^+e^-}(s) = \left( 1 + \epsrw \frac{s}{\mw^2 - s - i \mw \Gw} \right) F_\pi^V(s)
\eeq
(in line with $\hat t_R(s)_{12}$ in Eq.~\eqref{hattR}),
where $\epsrw$ will be determined by the fit.

In our formalism, the differential decay spectrum of $\eta' \to \pi^+ \pi^- \gamma$ is described by~\cite{Hanhart:2016pcd}
\begin{align}
\label{spectrum_pipigamma}
 \frac{\diff\Gamma(\eta'\to \pi^+\pi^-\gamma)}{\diff s}&=16\pi\alpha\Gamma_0|F_\pi^V(s)|^2 \bigg|P(s)\big(1+\Pi_\pi(s)\big)\notag\\
 - \frac{e^2 F_{\eta'\gamma\gamma}}{s}&-\frac{g_{\eta'\omega\gamma}}{\gwg}\frac{\epsrw-e^2\gwg^2}{\mw^2-s-i\mw\Gw}\bigg|^2,\notag\\
 \Gamma_0&=\frac{2s}{3}\bigg(\frac{M_{\eta'}^2-s}{16\pi M_{\eta'}}\sigma_\pi(s)\bigg)^3,
\end{align}
see \ref{app:decays+TFFs} for the derivation.
The appearance of the pion VFF herein is due to the universality of $\pi\pi$ $P$-wave final-state interactions~\cite{Stollenwerk:2011zz,Dai:2017tew}.
The constants $A$, $\beta$, and $\gamma$ in $P(s)$, see Eq.~\eqref{defP}, are used as fit parameters, and we refer to Eqs.~\eqref{Lagr_omega_gamma} and \eqref{gPVgamma} for the definitions of the coupling constants $\gwg$ and $g_{\eta'\omega\gamma}$, respectively. The pion VP $\Pi_\pi(s)$ is defined in Eq.~\eqref{Pi_pi}. Note that the residue of the $\omega$ propagator includes explicitly the one-photon-reducible contribution to $\rho$--$\omega$ mixing, $-e^2\gwg^2$, reflecting the fact that all form factors are defined excluding VP corrections. 

We fit to several time-like pion VFF data sets: provided by the $e^+ e^- \to \pi^+ \pi^-$ energy-scan experiments SND~\cite{Achasov:2005rg,Achasov:2006vp} and CMD-2~\cite{Akhmetshin:2001ig,Akhmetshin:2003zn,Akhmetshin:2006wh,Akhmetshin:2006bx}, where in both cases diagonal errors were given, as well as from radiative-return measurements BaBar~\cite{Aubert:2009ad,Lees:2012cj} (below $1 \GeV$) and KLOE~\cite{Ambrosino:2008aa,Ambrosino:2010bv,Babusci:2012rp,Anastasi:2017eio}, where in both cases statistical and systematic covariance matrices were provided. Furthermore, the recent data set for the $\eta' \to \pi^+ \pi^- \gamma$ spectrum measured by BESIII~\cite{BESIII:2017kyd} is used in the fit, which largely supersedes older data in statistical accuracy~\cite{CrystalBarrel:1997kku}. In order to avoid the d'Agostini bias~\cite{DAgostini:1993arp} in the minimization of a $\chi^2$ with naively constructed covariance matrices when dealing with normalization uncertainties, an iterative fit procedure, proposed in Ref.~\cite{Ball:2009qv} and applied to $e^+ e^- \to \pi^+ \pi^-$ in Ref.~\cite{Colangelo:2018mtw}, is employed. Since the pion VFF is defined as a pure QCD quantity in Eq.~\eqref{eq:pvff_def}, the pion VFF data sets have been undressed of VP effects. As noted in Ref.~\cite{Colangelo:2018mtw}, a minor rescaling of energy for each individual $\pi^+ \pi^-$ data set leads to a significant improvement of the fit quality. Equivalently, we take the $\omega$ mass for individual data sets as a fit parameter. In the case of fitting the combined KLOE data set, we follow Ref.~\cite{Colangelo:2018mtw} and assign a global $\omega$ mass with individual mass shifts $\Delta \mw^{(i)}$ to each of the three underlying data sets, from 2008~\cite{Ambrosino:2008aa}, 2010~\cite{Ambrosino:2010bv}, and 2012~\cite{Babusci:2012rp}. Furthermore, the mass shifts are constrained by penalties $\Delta \chi^2_i = (\Delta \mw^{(i)} / \Delta E_\text{c})^2$, with calibration uncertainty $\Delta E_\text{c} = 0.2 \MeV$~\cite{KLOE}. These terms are counted as additional data points in the number of degrees of freedom. Finally, for BaBar and KLOE the observables are cross sections weighted over energy bins, an effect that is also included in our fit in the same way as in Ref.~\cite{Colangelo:2018mtw}. Further data sets are available from BESIII~\cite{Ablikim:2015orh} and SND~\cite{SND:2020nwa}, but not yet included for consistency: since we use the $\pi\pi$ phase shift from Ref.~\cite{Colangelo:2018mtw}, we restrict our analysis to the same data sets used therein, in particular, since the results for the $\eta'$ decays are insensitive to the precise choice of VFF data sets.

\begin{table}[t]
\renewcommand{\arraystretch}{1.3}
	\begin{tabular}{llr}
		\toprule
		Quantity & Value & Ref.\\\midrule
		$\metap$ & $957.78(6)\MeV$ & \cite{ParticleDataGroup:2020ssz}\\
		$\Getap$ & $0.188(6)\MeV$ & \cite{ParticleDataGroup:2020ssz}\\
		$F_{\eta'\gamma\gamma}$ & $0.3437(55)\GeV^{-1}$ & \cite{ParticleDataGroup:2020ssz}\\
		$\Br(\eta' \to \pi^+ \pi^- \gamma)$ & $29.5(4) \%$ & \cite{ParticleDataGroup:2020ssz}\\
		$\mw$ & $782.607(23)\MeV$ & \cite{Hoid:2020xjs}\\
		$\Gw$ & $8.69(4)\MeV$ & \cite{Hoid:2020xjs}\\
		$\Gamma(\omega\to e^+e^-)\Br(\omega\to3\pi)$ & $0.5698(31)(82)\keV$ & \cite{BABAR:2021cde}\\
		$\Br(\omega\to3\pi)$ & $89.2(7)\%$ & \cite{ParticleDataGroup:2020ssz}\\
		$\Br(\eta' \to \omega \gamma)$ & $2.50(7) \%$ &\cite{ParticleDataGroup:2020ssz}\\
		$\mphi$ & $1019.197(20)\MeV$ & \cite{Hoid:2020xjs}\\
		$\Gphi$ & $4.22(5)\MeV$ & \cite{Hoid:2020xjs}\\
		$\Gamma(\phi\to e^+e^-)\Br(\phi\to3\pi)$ & $0.1841(21)(80)\keV$ & \cite{BABAR:2021cde}\\
		$\Br(\phi\to3\pi)$ & $15.24(33)\%$ & \cite{ParticleDataGroup:2020ssz}\\
		$\Br(\phi\to\eta' \gamma)$ & $6.22(21)\times 10^{-5}$ &\cite{ParticleDataGroup:2020ssz}\\
		\bottomrule
	\end{tabular}
	\centering
	\caption{Input parameters used in this work. Note that $\mw$ becomes a free parameter for the VFF fits, to account for the uncertainty in energy calibration in each data set as well as the tension with determinations from $e^+e^-\to 3\pi$ and $e^+e^-\to \pi^0\gamma$. The quoted numbers for $\mw$, $\Gw$ are VP subtracted, to ensure consistency with the bookkeeping as defined in the appendix. We also show the analog quantities for the $\phi$, which enter the isoscalar part of the TFF (and are consistent with analogous determinations from $e^+e^-\to\bar KK$~\cite{Stamen:2022uqh}). The entries for $\Gamma(V\to e^+e^-)\Br(V\to3\pi)$, $V=\omega,\phi$, from Ref.~\cite{BABAR:2021cde} are consistent with but more precise than the current PDG averages.}
	\label{tab:fit_input}
\renewcommand{\arraystretch}{1.0}
\end{table}

The BESIII data set for $\eta' \to \pi^+ \pi^- \gamma$ includes the number of selected events, the number of background events, the detection efficiency, and the detector resolution root mean square in the respective energy bin. The latter is taken into account by taking the convolution of the theoretical spectrum in Eq.~\eqref{spectrum_pipigamma} with a Gaussian distribution of mean corresponding to the respective bin center and standard deviation given by the respective value for the energy resolution. Efficiency and number of background events are subsequently included in the fit function. Since the data are given in the form of an event rate, the physical values of the fit parameters $A$ and $\epsrw$ are extracted by utilizing the constraint
\beq
    \Gamma(\eta' \to \pi^+ \pi^- \gamma) = \int_{4 \mpi^2}^{\metap^2} \diff s\, \frac{\diff \Gamma(\eta' \to \pi^+ \pi^- \gamma)}{\diff s}
\label{eq:int_diffwidth}
\eeq
in the single fit to the BESIII decay spectrum, where the partial width $\Gamma(\eta' \to \pi^+ \pi^- \gamma)$ is given by the corresponding quantities in Table~\ref{tab:fit_input}. The couplings $g_{\omega \gamma}$ and $g_{\eta' \omega \gamma}$ have been extracted by means of VMD models~\cite{Hanhart:2016pcd} to be
\begin{align}
\label{eq:couplings_gog_geog}
    g_{\omega \gamma} &= \sqrt{\frac{3 \Gamma(\omega \to e^+ e^-)}{4 \pi \alpha^2 \mw}} = 0.0619(3),\\
    g_{\eta' \omega \gamma} &= - \sqrt{\frac{8 \metap^3 \Gamma(\eta' \to \omega \gamma)}{\alpha(\metap^2 - \mw^2)^3}} = -0.400(8) \GeV^{-1},\notag
\end{align}
where the experimental input quantities are listed in Table \ref{tab:fit_input} as well. Note that the definition of the coupling $g_{\eta' \omega \gamma}$ differs from the one in Ref.~\cite{Hanhart:2016pcd} by a factor of the elementary charge $e$, with such factors always factorized throughout this work. Also note that the given value for the coupling $g_{\omega \gamma}$ has been corrected for VP in order to be consistent with the definition of the pion VFF as a pure QCD quantity. The correction amounts to a factor of $|1-\Pi(\mw^2)|=0.977$, where $\Pi(s)$ is the Standard Model VP function of Ref.~\cite{Keshavarzi:2018mgv}.

\begin{table}[t!]
\renewcommand{\arraystretch}{1.3}
	\begin{tabular}{lrr}
	    \toprule
		Quantity & Likelihood & $\chi^2$ \\ \midrule
		$A \ [\GeV^{-3}]$ & $17.12(35)$ & $17.09(32)$ \\
		$\beta \ [\GeV^{-2}]$ & $0.714(55)$ & $0.723(45)$ \\
		$\gamma \ [\GeV^{-4}]$ & $-0.412(55)$ & $-0.420(45)$\\
		$\epsrw \times 10^{3}$ & $1.998(67)$ & $1.997(54)$\\
		$\mw \ [\MeV]$ & $782.99(33)$ & $783.00(27)$\\
	\bottomrule
	\end{tabular}
    \centering
\caption{Comparison of the fit outcome of the differential decay width in Eq.~\eqref{spectrum_pipigamma} to the BESIII $\eta'\to \pi^+ \pi^- \gamma$ spectrum~\cite{BESIII:2017kyd} of the binned maximum likelihood and minimum $\chi^2$ strategies.}
\label{tab:likevschi}
\renewcommand{\arraystretch}{1.0}
\end{table}

In principle, the provided data would need to be fit by the binned maximum likelihood strategy, i.e., by minimizing the log-likelihood function
\beq
    \mathcal{L} = - \sum\limits_{i} \log \left( \frac{(\mu_i)^{n_i} e^{- \mu_i}}{n_i!} \right),
\eeq
where $n_i$ represents the number of events in bin $i$ and $\mu_i$ is the value of the fit function in this energy bin. This proves to be disadvantageous for our purpose, since we aim at a combined $\chi^2$ fit of the pion VFF data sets and the $\eta' \to \pi^+ \pi^- \gamma$ spectrum in order to extract a common $\rho$--$\omega$ mixing parameter $\epsrw$. In Table~\ref{tab:likevschi}, the fit parameters to the BESIII data stemming from the binned maximum likelihood method are compared to those of a $\chi^2$ fit, by minimizing
\beq
    \chi^2 = \sum\limits_{i} \left(\frac{\mu_i - n_i}{\sigma_i}\right)^2,
\eeq 
where Poissonian errors $\sigma_i = \sqrt{n_i}$ have been assumed. The $\chi^2$ fit gives a $\chi^2/\text{dof} = 1.30$ and the errors of the fit parameters have been inflated by a scale factor $\sqrt{\chi^2/\text{dof}}$. While the central values all agree with each other within error margins, the error estimates of the $\chi^2$ fit are slightly lower than those of the binned maximum likelihood fit, but sufficiently close to justify a $\chi^2$ fit in combination with VFF data. The quoted errors for the extracted values of $A$ and $\epsrw$ include the errors of all fit parameters and their correlations through Eq.~\eqref{eq:int_diffwidth} as well as the errors of the experimental input parameters.

\begin{table*}[t!]
\centering
\renewcommand{\arraystretch}{1.3}
	\begin{tabular}{llrlllrr}
	   \toprule
		& $\chi^2/\text{dof}$ & $\mw \ [\MeV]$ & $A \ [\GeV^{-3}]$ & $\beta \ [\GeV^{-2}]$ & $\gamma \ [\GeV^{-4}]$ & $\alpha_\pi \times 10^2 \ [\GeV^{-2}]$ & $\epsrw \times 10^3$ \\ \midrule
		BaBar & $1.26$ & $781.875(82)$ & \multirow{5}{*}{\phantom{.}} & \multirow{4}{*}{\phantom{.}} & \multirow{5}{*}{\phantom{.}} & \multirow{7}{*}{$\left . \vphantom{\begin{tabular}{r} a\\ a\\ a\\ a\\ a\\ a\\a\\ \end{tabular}} \right\rbrace 5.74(14)$} & \multirow{7}{*}{$\left . \vphantom{\begin{tabular}{r} a\\ a\\ a\\ a\\ a\\ a\\a\\ \end{tabular}} \right\rbrace 2.007(10)$}\\
		KLOE & $1.61$ & $\left\lbrace \begin{array}{r} 781.65(12) \\ 782.10(17)
		\\ 781.84(27)\end{array}\right.$ & & & & & \\
		CMD-2 & $2.18$ & $782.131(68)$ & & & & &\\
		SND & $2.16$ & $781.457(97)$ & & & & &\\
		BESIII & $1.31$ & $783.00(28)$ & $17.10(32)$ & $0.720(46)$ & $-0.418(46)$ & &\\ \bottomrule
	\end{tabular}
    \caption{Combined fit to several pion VFF data sets (BaBar, KLOE, CMD-2, SND) and $\eta' \to \pi^+ \pi^- \gamma$ spectrum (BESIII) with overall $\chi^2/\text{dof} = 1.46$. In the row for KLOE, the three values for $\mw$ refer to the combinations of the global KLOE $\omega$ mass and the corresponding mass shifts of the three underlying data sets from 2008, 2010, 2012, respectively. See main text for details.}
    \label{tab:fitcombined}
    \renewcommand{\arraystretch}{1.0}
\end{table*}

In the combined fit of the $\eta' \to \pi^+ \pi^- \gamma$ spectrum and the pion VFF data sets, the $\rho$--$\omega$ mixing parameter $\epsrw$ is a shared parameter. In comparison to the $\eta' \to \pi^+ \pi^- \gamma$ single fit, we are confronted with the problem that we fit the physical value of $\epsrw$ and not the one including the event rate $N_\text{ev}$. Therefore, $N_\text{ev}$ needs to be calculated at every step of the fit iteration. Defining $2P_\text{ev}(s) = N_\text{ev} A(1 + \beta s + \gamma s^2)$, Eq.~\eqref{eq:int_diffwidth} can be written as
\beq
    0 = \mathcal{A}_2 N_\text{ev}^2 + \mathcal{A}_1 N_\text{ev} + \mathcal{A}_0,
\eeq
where
\begin{align}
\label{fit_coefficients}
    \mathcal{A}_2 &= - \Gamma(\eta' \to \pi^+ \pi^- \gamma) + 16 \pi \alpha \int_{4\mpi^2}^{\metap^2} \diff s\, \Gamma_0 |F_\pi^V(s)|^2\notag \\
    &\qquad \times \left| \frac{g_{\eta' \omega\gamma}}{\gwg}\frac{\epsrw-e^2\gwg^2}{\mw^2-s-i\mw\Gw} + \frac{e^2 F_{\eta' \gamma\gamma}}{s} \right|^2,\notag \\ 
    \mathcal{A}_1 &= 32 \pi \alpha \int_{4\mpi^2}^{\metap^2} \diff s\, \Gamma_0 |F_\pi^V(s)|^2\ \mathrm{Re} \bigg[ P_\text{ev}(s)\big(1 + \Pi_\pi^*(s)\big)\notag \\ 
    &\qquad\times \bigg(\frac{g_{\eta' \omega\gamma}}{\gwg}\frac{e^2\gwg^2 - \epsrw}{\mw^2-s-i\mw\Gw} - \frac{e^2 F_{\eta' \gamma\gamma}}{s} \bigg) \bigg], \notag \\ 
    \mathcal{A}_0&= 16 \pi \alpha \int_{4\mpi^2}^{\metap^2} \diff s\, \Gamma_0 |F_\pi^V(s)|^2 P_\text{ev}^2(s) \big|1 + \Pi_\pi(s) \big|^2.
\end{align}
and notation as in Eq.~\eqref{spectrum_pipigamma},
in order to determine $N_\text{ev}$ with input of $\epsrw$ in physical units. Hence, the main difference to the individual fit is that to be able to perform a combined fit with the VFF data sets, we need to ensure physical units already at each step in the fit iteration. The linear parameter $\alpha_\pi$ in the pion VFF is used as a shared parameter for all data sets in the combined fit as well. The pion VP appearing in the $\eta'$ decay amplitude is included in a self-consistent way.

\begin{figure}[t!]
    \centering
    \includegraphics[width=\linewidth]{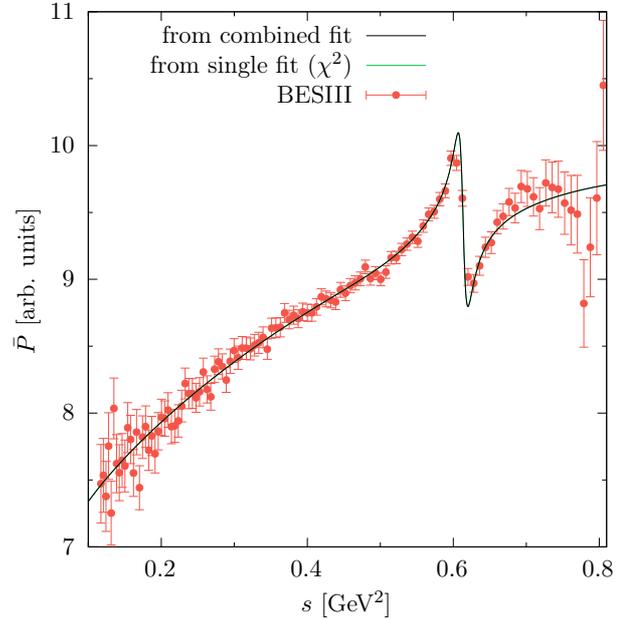}
    \caption{Fit to the differential decay rate of $\eta' \to \pi^+ \pi^- \gamma$ (individually or combined with the VFF). To highlight potential differences in the $\rho$--$\omega$ region, we show the associated function $\bar{P}$, as defined in Eq.~\eqref{eq:barP}, compared to the experimental data from BESIII~\cite{BESIII:2017kyd}. The function $\bar{P}$ is calculated in units that still contain the event rate. The two fits cannot be distinguished on this scale.}
    \label{fig:etappg_spec}
\end{figure}

The result of the combined fit is listed in Table~\ref{tab:fitcombined}, where the errors are inflated by the overall $\sqrt{\chi^2/\text{dof}}$. The error of the physical value of $A$ takes into account the correlations and inflated errors of the fit parameters as well as the errors of all input quantities. The results for $\mw$ and $\epsrw$ are in good agreement with Ref.~\cite{Colangelo:2018mtw}, validating the simplified fit form~\eqref{eq:pvff_omnes} as a convenient way to compare $\rho$--$\omega$ mixing in the pion VFF and $\eta'\to\pi^+\pi^-\gamma$. The outcome of the combined fit for the case of the $\eta' \to \pi^+ \pi^- \gamma$ spectrum is shown in Fig.~\ref{fig:etappg_spec}, where
\beq
\label{eq:barP}
\bar{P}(s) = \left[\frac{1}{\Gamma_0 |\Omega(s)|^2}\frac{\diff \Gamma(\eta' \to \pi^+ \pi^- \gamma)}{\diff s} \right]^{1/2}
\eeq
is shown to emphasize potential differences in the vicinity of the $\rho$--$\omega$ mixing effect.

\nocite{Abbiendi:2022liz,Colangelo:2022prz,BESIII:2019gjz,Achasov:2003ir}

\begin{figure*}[t!]
    \begin{subfigure}{0.49\textwidth}
    \includegraphics[width=\linewidth]{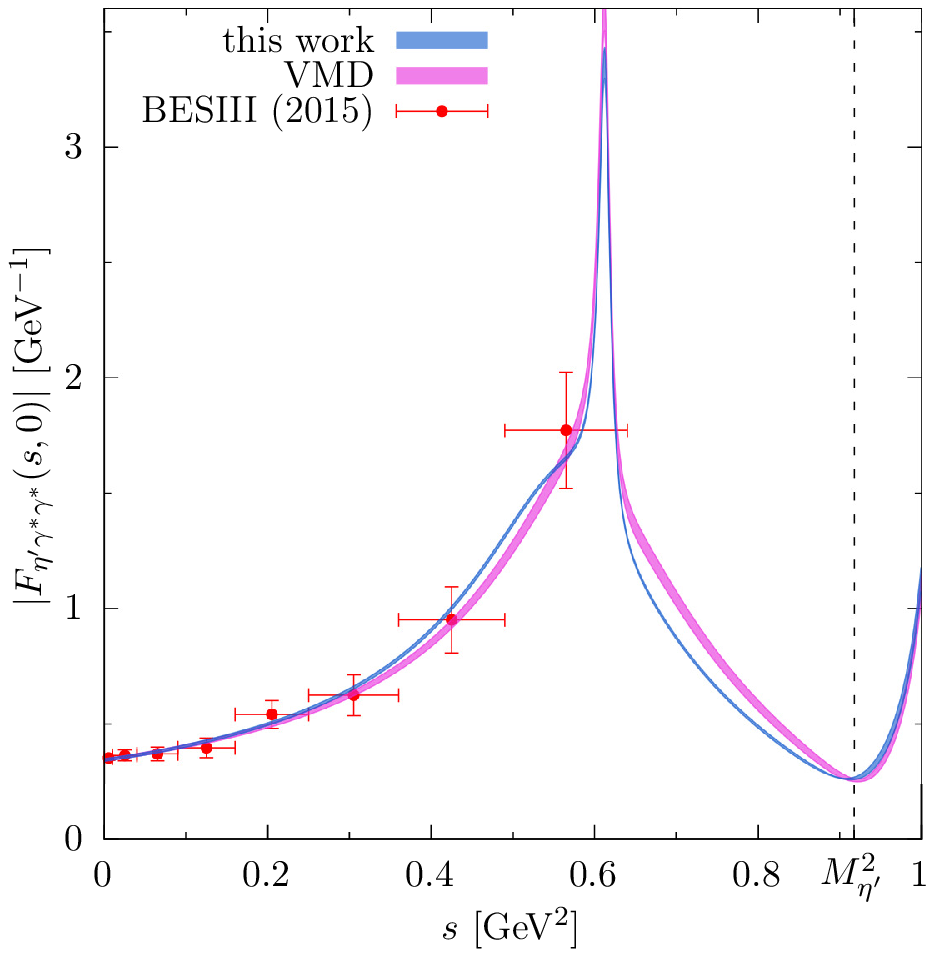}
    \end{subfigure}
    \hfill
     \begin{subfigure}{0.49\textwidth}
    \includegraphics[width=\linewidth]{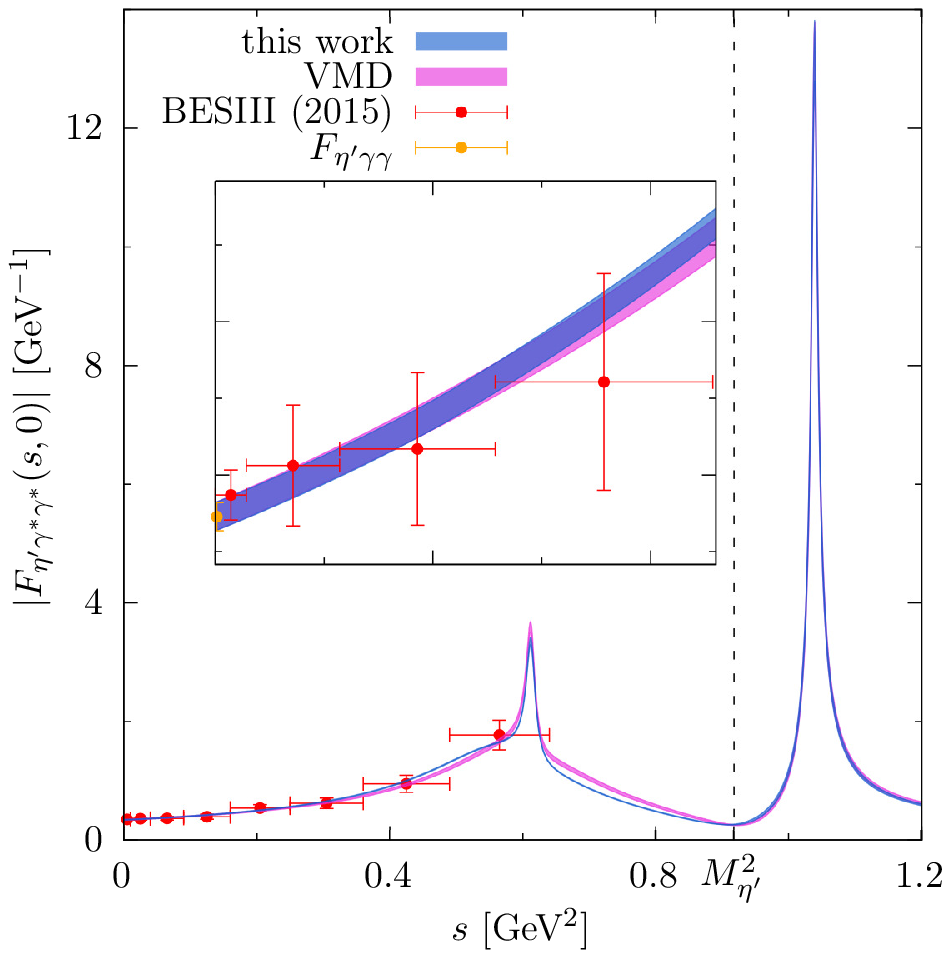}
    \end{subfigure}
    \hfill
    \caption{Determination of the $\eta'$ TFF by means of Eq.~\eqref{rep_final} in comparison to data from BESIII~\cite{BESIII:2015zpz} (statistical and systematic errors added in quadrature) scaled with $F_{\eta' \gamma \gamma}$ from Table~\ref{tab:fit_input} and the VMD model of Eq.~\eqref{rep_final_VMD} with $\phi$ resonance contribution according to Eq.~\eqref{eq:phi_contribution}; for the kinematic range accessible in $\eta'$ decays (left) and a larger time-like region including the $\phi$ resonance with inset magnifying the low-$s$ region (right). See also Ref.~\cite{Holz:2016} for an earlier version of this figure.}
    \label{fig:etaptff}
\end{figure*}

While Fig.~\ref{fig:etappg_spec} and the comparison of 
 Tables~\ref{tab:likevschi} and~\ref{tab:fitcombined}
 shows that both processes are fully consistent regarding $\epsrw$, 
 a minor tension occurs in the mass of the $\omega$.\footnote{Note that constraining the relative shifts in the $\omega$ masses from the expected uncertainty in the energy calibration, in a similar fashion as it was done here in the combined fit for the case of KLOE, only leads to marginal changes in the  results compared to the ones displayed in Table~\ref{tab:fitcombined} for the other pion VFF data sets, see Ref.~\cite{Colangelo:2018mtw}.} Here, the results from $2\pi$ come out significantly below the preferred value from $e^+e^-\to 3\pi, \pi^0\gamma$~\cite{Hoferichter:2019mqg,Hoid:2020xjs}, see Table~\ref{tab:fit_input}, but extractions from $2\pi$ are known to be sensitive to a phase in $\epsrw$~\cite{Lees:2012cj}, with the corresponding difference at least partially explained by radiative intermediate states that can generate such an isospin-breaking effect~\cite{Abbiendi:2022liz,Colangelo:2022prz}. In contrast, the $\eta'\to\pi^+\pi^-\gamma$ data favor a central value that deviates about $0.5\MeV$ in the opposite direction, albeit with moderate significance. Since also the recent $3\pi$ data from BESIII~\cite{BESIII:2019gjz} suggest a similar increase compared, e.g., to Refs.~\cite{Achasov:2003ir,Akhmetshin:2003zn,BABAR:2021cde}, as does Ref.~\cite{Ablikim:2015orh} compared to other $2\pi$ data sets, it is difficult not to see a pattern in BESIII determinations of $\mw$.   

\section{Predicting $\eta'\to\ell^+\ell^-\gamma$}
\label{sec:etap_llg}

In order to determine the $\eta'$ TFF according to Eq.~\eqref{rep_final}, the $\rho$--$\omega$ mixing parameter $\epsrw$, the parameter associated with inelastic contributions to the pion VFF $\alpha_\pi$, as well as the polynomial parameters $A$, $\beta$, and $\gamma$ of the $\eta' \to \pi^+ \pi^- \gamma$ spectrum from the combined fit serve as input. Additionally, from the quantities listed in Table~\ref{tab:fit_input}, the $\omega$ mass and width, the TFF normalization $F_{\eta' \gamma \gamma}$, together with the associated couplings in Eq.~\eqref{eq:couplings_gog_geog} enter the description of the isovector part of the TFF. The isoscalar part, consisting of the contributions of the $\omega$ and $\phi$ resonances, depends on the couplings $w_{\eta' V \gamma}$ in Eq.~\eqref{eq:isoscalar_coup}, which were calculated from the quantities in Table~\ref{tab:fit_input}. In the time-like regime, the resulting TFF then appears as shown in Fig.~\ref{fig:etaptff}, where it is compared to experimental data~\cite{BESIII:2015zpz}.

\begin{table*}[t!]
    \renewcommand{\arraystretch}{1.3}
    \centering
    \begin{tabular}{lllllll}
       \toprule
		& $(I=1)_{\epsrw=0}$ & $\Delta(I=1)_{\epsrw \neq 0}$ & $(I=0)^\omega_{\epsrw=0}$ & $\Delta (I=0)^\omega_{\epsrw \neq 0}$ & $(I=0)^\phi$ & Total\\ \midrule
		norm $[\%]$ & $69.18(86)$ & $-0.1388(19)$ & $7.06(22)$ & $-0.1397(47)$ & $15.85(61)$ & $91.9(1.1)$ \\
		$b_{\eta'} \ [\GeV^{-2}]$ & $1.160(23)$ & $0$ & $0.1176(32)$ & $0$ & $0.1526(53)$ & $1.431(23)$\\ \bottomrule
    \end{tabular}
    \caption{Contributions from the various components of the TFF to the sum rules of the normalization and the slope parameter.}
    \label{tab:norm_slope}
    \renewcommand{\arraystretch}{1.0}
\end{table*}

Since the form of polynomial $P(s)$ and the linear polynomial multiplying the Omn\`es function in the representation of the pion VFF are governed by physics below $1 \GeV$, their values at high energies do not bear much meaning. Therefore, and in order to improve the convergence properties of the dispersion integrals in Eq.~\eqref{rep_final} at the same time, they are led to constant values $P(s_c)$ and $1 + \alpha_\pi s_c$ above a certain cutoff value $s_c$. This cutoff is varied between $s_c=1 \GeV^2$, the point where $P(s)$ reaches its maximum ($s_c = 0.86 \GeV^2$ with $\beta$ and $\gamma$ from the combined fit) and the point where $P$ drops below its value at $s=0$ ($s_c = 1.72 \GeV^2$). Together with the errors of the input quantities in Table~\ref{tab:fit_input}, the errors of the parameters from the combined fit, and their correlations, this procedure is used to generate the error band of the dispersive  $\eta'$ TFF representation shown in Fig.~\ref{fig:etaptff}. In this figure, the dispersive curve displays the broad $\rho$ peak from the isovector part of the TFF. Around the $\omega$ mass, the $\rho$--$\omega$ mixing effect is overlaid by the narrow peak of the $\omega$ resonance from the isoscalar part of the TFF. Slightly below an invariant mass square of $1 \GeV^2$ the isoscalar $\phi$ contribution begins to set in. The available experimental data are in excellent agreement with the dispersive prediction, but not accurate enough to distinguish it from the VMD model. Figure~\ref{fig:etaptff} also shows the peak of the $\phi$ resonance, which, however, is not accessible in $\eta'$ decays, but could be scanned in $e^+e^-\to\eta'\gamma$ near threshold.

The low-energy properties of the TFF are described by 
the slope parameter $b_{\eta'}$, which is defined by
\beq
    F_{\eta' \gamma^* \gamma^*}(s,0) = F_{\eta' \gamma \gamma} \big(1 + b_{\eta'} s + \Order\big(s^2\big)\big).
\eeq
It can be calculated via a sum rule, following from Eq.~\eqref{rep_final}, and disentangled into its isovector contribution
\beq
    b_{\eta'}^{(I=1)} = \frac{1}{48 \pi^2 F_{\eta' \gamma \gamma}} \int_{4\mpi^2}^{\infty} \frac{\diff x}{x}\, \sigma^3(x) P(x) |F_\pi^V(x)|^2,
\eeq
and isoscalar contribution
\beq
    b_{\eta'}^{(I=0)} = \sum\limits_{V \in \{\omega, \phi\}} \frac{w_{\eta' V \gamma}}{M_V^2},
\eeq 
with $b_{\eta'} = b_{\eta'}^{(I=1)}  + b_{\eta'}^{(I=0)}$. Numerical values are listed together with the evaluation of the sum rule for the normalization~\eqref{sum_rule} in Table~\ref{tab:norm_slope}. As expected from VMD, the isovector part dominates both the normalization and the slope, saturating the former by $70\%$ and producing about $80\%$ of the latter. In combination  with the isoscalar part the sum rule for the normalization is fulfilled by $92\%$, suggesting a slightly faster convergence than similar sum rules in Refs.~\cite{Schneider:2012ez,Hoferichter:2012pm,Hoferichter:2018dmo,Hoferichter:2018kwz}. The two isospin-breaking corrections to the normalization, in the isovector and isoscalar part, combine to about $-0.3\%$ and thus prove negligible compared to the uncertainty in the sum-rule evaluation. Accordingly, these corrections become most relevant in the resonance region, where even subleading effects are enhanced by the small width of the $\omega$ resonance, see Fig.~\ref{fig:etaptff}. The $\omega$ resonance peak would be about $4\%$ increased in magnitude if $\epsrw=0$.

Our result for the slope,
\beq
b_{\eta'}=1.431(23)\GeV^{-2},
\eeq
is consistent with, but considerably more precise than the experimental determinations $b_{\eta'}=1.60(17)(8)\GeV^{-2}$~\cite{BESIII:2015zpz} (from $\eta'\to e^+e^-\gamma$) and $b_{\eta'}=1.60(16)\GeV^{-2}$~\cite{CELLO:1990klc}, $b_{\eta'}=1.38(23)\GeV^{-2}$~\cite{TPCTwoGamma:1990dho} (via $e^+e^-\to e^+e^- \eta'$), and also agrees very well with $b_{\eta'}=1.43(4)(1) \GeV^{-2}$~\cite{Escribano:2015yup} extracted via Pad\'e approximants. Moreover, the improved formalism and combined fit allowed us to substantially reduce the uncertainties compared to the previous dispersion-theoretical analysis $b_{\eta'}=1.53^{+0.15}_{-0.08}\GeV^{-2}$~\cite{Hanhart:2013vba}, also reflecting the improved input on $\eta'\to\pi^+\pi^-\gamma$ from Ref.~\cite{BESIII:2017kyd} in comparison to the earlier measurement~\cite{CrystalBarrel:1997kku}. A close-up of the low-$s$ region of the TFF is included in Fig.~\ref{fig:etaptff} as well.

\section{Conclusions}
\label{sec:conclusions}

This work presents progress in a dispersive calculation of the TFF of the $\eta'$, as required for future studies of the $\eta'$-pole contribution to HLbL scattering in the anomalous magnetic moment of the muon. First, we established a formalism that enables a consistent implementation of isospin-breaking $\rho$--$\omega$ mixing effects, both in the isoscalar and isovector part of the form factor. The technical derivation is presented in a self-contained way in the appendices of the paper, leading to our main result~\eqref{rep_final} for the final TFF representation. Moreover, we worked out how the information on the isovector TFF contained in the spectrum for $\eta'\to\pi^+\pi^-\gamma$ can be combined, in terms of the decomposition~\eqref{spectrum_pipigamma}.

These results form the basis for the application studied in the main part of the paper, an analysis of recent data from BESIII on $\eta'\to\pi^+\pi^-\gamma$ and subsequent prediction of the full $\eta'$ TFF.
The combined fit 
of the $\eta'\to\pi^+\pi^-\gamma$ spectrum and data on the pion VFF gives consistent results for  
the $\rho$--$\omega$ mixing parameter,  see Fig.~\ref{fig:etappg_spec}, but reveals a minor tension in  the $\omega$ mass. 
Given the limited statistics, it would be desirable to improve the fit with an independent measurement of the $\eta'\to\pi^+\pi^-\gamma$ spectrum, e.g., from CLAS at Jefferson Lab~\cite{MbiandaNjencheu:2017vwy}.  

As a final step, we used the parameters determined in the global fit to $\eta'\to\pi^+\pi^-\gamma$ and $e^+e^-\to\pi^+\pi^-$, in combination with up-to-date input on the isoscalar resonances, to predict the full TFF at low energies, see Fig.~\ref{fig:etaptff} and Table~\ref{tab:norm_slope}. Deviations from VMD are most visible in the vicinity of the $\omega$ resonance, where the dispersive implementation of the $\rho$ meson via $2\pi$ intermediate states changes the line shape compared to the strict VMD limit. Our results should prove valuable for the analysis of future data on $\eta'\to\pi^+\pi^-\gamma$ and $\eta'\to\ell^+\ell^-\gamma$~\cite{Heijkenskjold:2019hib,Heijkenskjold:2019xdi}, in particular towards improving the calculation of the $\eta'$-pole contribution to HLbL scattering.

\begin{acknowledgements}
We thank Andrzej Kup\'s\'c for heroic efforts to make the data from Ref.~\cite{BESIII:2017kyd} available to us, and Pablo
S\'anchez-Puertas for pointing out the issue of one-photon-reducible contributions to $\epsrw$. 
Financial support by the SNSF (Project Nos.\ PCEFP2\_181117 and 200020\_200553),
the DFG through the funds provided to the Sino--German Collaborative
Research Center TRR110 ``Symmetries and the Emergence of Structure in QCD'' (DFG Project-ID 196253076 -- TRR 110),
the Bonn--Cologne Graduate School of Physics and Astronomy (BCGS), the  US DOE (Grant
No.\ DE-FG02-00ER41132),
and the European Union's Horizon 2020 research and innovation programme under grant agreement No.\ 824093
is gratefully acknowledged.
\end{acknowledgements}

\appendix

\section{Coupled-channel formalism}
\label{app:coupled}

In~\ref{app:2piellell}--\ref{app:full}, we derive a coupled-channel formalism for $\pi^+\pi^-$, $\ell^+\ell^-$, and $\pi^+\pi^-\pi^0$, concentrating on the $\omega$ resonance for the latter. In particular, we aim at delineating the impact of $\rho$--$\omega$ mixing on the pion vector form factor and pseudoscalar decays in a consistent way, leading to the master formulae with which the $\eta'$ decays are analyzed in the main part. The construction follows the formalism developed in Ref.~\cite{Hanhart:2012wi} (cf.\ also Refs.~\cite{Ropertz:2018stk,VonDetten:2021rax}), which can be summarized as follows: the full scattering amplitude $t(s)$, carrying channel indices $i$, $j$, is written as\footnote{Note that, here and below, repeated indices related to $\xi_i$ and $\Gamma_i$ are not summed over.}
\beq
\label{t_decomp}
t(s)_{ij}=\tilde t(s)_{ij} + \xi_i \Gamma_\text{out}(s)_{i}t_R(s)_{ij}\Gamma_\text{in}(s)^\dagger_j \xi_j,
\eeq
where $t_R$ denotes the part of the scattering amplitude that arises from iterating a resonance potential
\beq
t_R(s)_{ij}=\big[\unity-V_R(s)\Sigma(s)\big]^{-1}_{ik}V_R(s)_{kj},
\label{tR_def}
\eeq
via the self energies
\beq
\Sigma(s)_{ij}=\delta_{ij}\frac{s}{\pi}\int_{s_\text{thr}^i}^\infty\diff s' \frac{\tilde \sigma_i(s')\xi_i^2|\Gamma(s')_i|^2}{s'(s'-s-i\eps)},
\eeq
with phase space $\tilde \sigma_i$, centrifugal barrier factors $\xi_i$, and vertex functions $\Gamma_i$. The potential is parameterized in terms of bare resonances 
\begin{align}
\bar V_R(s)_{ij}&=-\sum _{l,l'=1}^n g_i^{(l)}G_{ll'}g_j^{(l')}, \qquad G_{ll'}=\frac{\delta_{ll'}}{s-m_l^2},\notag\\
V_R(s)&=\bar V_R(s)-\bar V_R(0),
\end{align}
with bare masses $m_l$ and then (potentially) subtracted at $s=0$. For our application, this formalism is attractive, as it allows us to account for the physics of photon and $\omega$ exchange via $V_R$, the $\pi\pi$ rescattering via Omn\`es factors in the $\Gamma_i$, and combine everything in a way that is consistent with analyticity and unitarity. In addition, form factors (and thereby pseudoscalar decays) can be described in a similar way
\beq
\label{def_Fi}
F(s)_i=\Gamma_\text{out}(s)_i\big[\unity-V_R(s)\Sigma(s)\big]^{-1}_{ik}M_k,
\eeq
with source terms parameterized as
\beq
M_k=c_k-\sum_{l,l'=1}^ng_k^{(l)}G_{ll'}\alpha^{(l')}s,
\label{MK_ansatz}
\eeq
where the factor $s$ in the resonance coupling arises as a consequence of gauge invariance. In the following, we will develop this formalism step by step for a three-channel system $\{2\pi,\ell^+\ell^-,3\pi\}$, mediated via two resonances $\{\gamma,\omega\}$. 
Note that the $\rho$ resonance is already contained in $\tilde t(s)$ for the $2\pi$ channel and thus also in the corresponding vertex function $\Gamma(s)$.

\section{$\pi^+\pi^-$ and $\ell^+\ell^-$}
\label{app:2piellell}

\subsection{$\pi\pi$ channel}

As a first step we consider the single-channel case when only external $\pi^+\pi^-$ states and photon exchange are retained. The isospin $I=1$ $\pi\pi$ scattering amplitude is written in the conventions
\beq
\M^{1}(s,t)=32\pi\sum_{\text{odd }l}(2l+1)t^1_l(s)P_l(z),
\eeq
with Legendre polynomials $P_l(z)$ and normalized in such a way that the cross section reads
\beq
\frac{\diff \sigma}{\diff\Omega}=\frac{|\M|^2}{64\pi^2 s}.
\eeq
The $P$-wave amplitude is identified with $\tilde t(s)$ in Eq.~\eqref{t_decomp} as
\begin{align}
\label{t11}
\tilde t(s)&\equiv 48\pi t^1_1(s)=\frac{e^{2i\delta^1_1(s)}-1}{2i\tilde \sigma_\pi(s)},\notag\\
\tilde \sigma_\pi(s)&=\frac{\sigma_\pi(s)}{48\pi},\qquad 
\sigma_\pi(s)=\sqrt{1-\frac{4\mpi^2}{s}}. 
\end{align}
This definition is motivated by the isospin decomposition of the physical amplitude
\beq
\M^{\pi^+\pi^-}=\frac{1}{6}\M^2-\frac{1}{2}\M^1+\frac{1}{3}\M^0,
\eeq
since then $\M^{\pi^+\pi^-}|_{P\text{-wave}}=-t(s) z$ and $t(s)$ can essentially be interpreted as the standard amplitude with the angular dependence removed.   

In addition, we can match the contribution from the $s$-channel photon-exchange diagram 
\beq
\label{pipi_photon_exchange}
\M^{\pi^+\pi^-}_\gamma=\frac{e^2}{s}\big(s-4\mpi^2\big)z\big(F_\pi^V(s)\big)^2
\eeq
to the second term in Eq.~\eqref{t_decomp}. That is, including the hadronic running of $\alpha$ for self consistency, we have
\begin{align}
t_\gamma(s)&=-\frac{4\pi\alpha}{s}\big(s-4\mpi^2\big)\big(F_\pi^V(s)\big)^2(1-\Pi_\pi(s))^{-1},\notag\\
\Pi_\pi(s)&=-\frac{4\pi\alpha}{s}\frac{s^2}{\pi}\int_{4\mpi^2}^\infty\diff s'\frac{\tilde \sigma_\pi(s')(s'-4\mpi^2)|F_\pi^V(s')|^2}{s'^2(s'-s-i\eps)},
\label{Pi_pi}
\end{align}
which takes the form
\begin{align}
t_\gamma(s)&=\frac{V_R(s)\xi_\pi^2(s)\Gamma_\text{out}(s)\Gamma_\text{in}^\dagger(s)}{1-V_R(s)\Sigma_\pi(s)}\notag\\
&=\xi_\pi^2(s)\Gamma_\text{out}(s)\Gamma_\text{in}^\dagger(s)t_R(s),
\end{align}
provided that we identify
\begin{align}
 \Gamma_\text{out}(s)&=\Gamma_\text{in}^\dagger(s)=F_\pi^V(s),\qquad V_R(s)=-\frac{4\pi\alpha}{s},\notag\\
 \xi_\pi(s)&=\sqrt{s-4\mpi^2},\notag\\
 \Sigma_\pi(s)&=\frac{s^2}{\pi}\int_{4\mpi^2}^\infty\diff s'\frac{\tilde\sigma_\pi(s')\xi_\pi^2(s')|\Gamma(s')|^2}{s'^2(s'-s-i\eps)}.
\end{align}
The full result
\beq
t(s)=\tilde t(s)+\xi_\pi(s)\Gamma_\text{out}(s)t_R(s)\Gamma_\text{in}^\dagger(s)\xi_\pi(s)
\eeq
then indeed agrees with Eq.~\eqref{t_decomp}, the only difference being that the self energy enters in twice-subtracted form. 

\subsection{$\ell^+\ell^-$ channel}

In the leptonic channel we start from the VP function in the form
\beq
\Pi_\ell(s)=-\frac{4\pi\alpha}{s}\frac{s^2}{\pi}\int_{4m_\ell^2}^\infty\diff s'\frac{\tilde\sigma_\ell(s')4(s'+2m_\ell^2)}{s'^2(s'-s-i\eps)},
\eeq
which leads one to identify
\begin{align}
\Gamma_\text{out}(s)&=\Gamma_\text{in}(s)=1,\qquad
V_R(s)=-\frac{4\pi\alpha}{s},\notag\\
\xi_\ell(s)&=2\sqrt{s+2m_\ell^2},\notag\\
\Sigma_\ell(s)&=\frac{s^2}{\pi}\int_{4m_\ell^2}^\infty\diff s'\frac{\tilde\sigma_\ell(s')\xi_\ell^2(s')}{s'^2(s'-s-i\eps)},
\end{align}
and thereby
\beq
t(s)=\xi_\ell^2(s) t_R(s)=\frac{\xi_\ell^2(s)V_R(s)}{1-V_R(s)\Sigma_\ell(s)}.
\eeq
This amplitude, when interpreted in the same way as the $\pi\pi$ amplitude $t(s)$ derived in the previous subsection, 
produces the total cross section\footnote{The factor $2/3$ arises from the angular integral $\int_{-1}^1\diff z\, z^2$, as does represent the correct angular dependence for the bosonic amplitude in Eq.~\eqref{pipi_photon_exchange}, but in general only the integrated quantities are reproduced correctly, since part of the angular dependence is moved into the centrifugal barrier factors, e.g., in contrast to $\xi_\pi$, the $\xi_\ell$ do not correspond to the center-of-mass momentum. Accordingly, we only keep the differential form when adequate, and otherwise simply display the integrated result.}
\begin{align}
\label{e+e-_check}
\sigma(\ell^+\ell^-\to\ell^+\ell^-)
&=\frac{2\pi}{64\pi^2 s}\frac{1}{4}\Big|\xi_\ell^2(s) V_R(s)\Big|^2 \times \frac{2}{3}\notag\\
&=\frac{(4\pi\alpha)^2}{12\pi s^3}(s+2m_\ell^2)^2,
\end{align}
which matches the QED result from the $s$-channel diagram (the factor $1/4$ averages over initial-state spins). In this way, the factor $\xi_\ell$ captures the kinematic dependence from the fermion traces instead of the centrifugal barrier, but up to this difference in interpretation the same decomposition applies as in the $\pi\pi$ channel. 

\subsection{Combining $\pi^+\pi^-$ and $\ell^+\ell^-$}

The two-channel system $\{2\pi,\ell^+\ell^-\}$ with a photon mediator in $V_R(s)$ follows by combining the results from the previous two subsections: 
\begin{align}
\label{t_R_def}
t_R(s)&=\big(\unity -V_R(s)\Sigma(s)\big)^{-1}V_R(s),\\
V_R(s)&=-\frac{4\pi\alpha}{s}\begin{pmatrix} 1 & 1\\ 1 & 1 \\ \end{pmatrix},
\qquad\! \Sigma(s)=\text{diag}\big(\Sigma_\pi(s),\Sigma_\ell(s)\big),\notag
\end{align}
so that
\begin{align}
t_R(s)&=-\frac{4\pi\alpha}{s}\begin{pmatrix} 1 & 1\\ 1 & 1 \\ \end{pmatrix}\big(1-\Pi(s)\big)^{-1},\notag\\
\Pi(s)&=-\frac{4\pi\alpha}{s}\big(\Sigma_\pi(s)+\Sigma_\ell(s)\big).
\end{align}
In particular, we find for the $e^+e^-\to\pi^+\pi^-$ cross section, by the same procedure as in Eq.~\eqref{e+e-_check}, 
\begin{align}
\label{pi+pi-_check}
\sigma(e^+e^-\to\pi^+\pi^-)&=\frac{2\pi}{64\pi^2 s}\frac{1}{4}\frac{\sigma_\pi(s)}{\sigma_e(s)} \big|t(s)_{21}\big|^2\times\frac{2}{3}\notag\\
&=\frac{\pi\alpha^2}{3s}\frac{\sigma_\pi^3(s)|F_\pi^V(s)|^2}{|1-\Pi(s)|^2}\frac{s+2\me^2}{s\sigma_e(s)},
\end{align}
which reproduces the expected result. Adding further leptonic channels is straightforward, e.g., in the VP function the sum of the lepton species appears. For simplicity, we will restrict the discussion to a single lepton channel throughout. 

\section{Including the $3\pi$ channel}
\label{app:3pi}

Our main objective in including the $3\pi$ channel is as follows: since in the end we are mainly interested in the $2\pi$ channel, the impact of $3\pi$ states is only indirect, so that we do not aim at a full description of the $\gamma^*\to3\pi$ amplitude as derived in Refs.~\cite{Hoferichter:2014vra,Hoferichter:2018kwz,Hoferichter:2018dmo}, but at a consistent implementation of the $\omega$, which, due to $\rho$--$\omega$ mixing, is how the $3\pi$ channel leaves its imprints in $2\pi$. This motivates including the $\omega$ as a second mediator besides the photon in the resonance potential $V_R$. 

\subsection{$3\pi$ and $\ell^+\ell^-$}

As a first step, we consider the $\ell^+\ell^-$ and $3\pi$ channels only, in order to fix the parameters in the potential. We make the ansatz
\begin{align}
\label{ansatz}
V_R(s)&=-\frac{4\pi\alpha}{s}\begin{pmatrix} 1 & g_3 s\\ g_3 s & (g_3 s)^2 \\ \end{pmatrix} - \frac{1}{s-\mwbare^2}\begin{pmatrix} 0 & 0\\ 0 & \gw^2 \\ \end{pmatrix},\notag\\
\Sigma(s)&=\text{diag}\big(\Sigma_\ell(s),\Sw(s)\big),
\end{align}
with functions/couplings $g_3$, $g_{\omega3}$, $\Sw(s)$, and bare $\omega$ mass $\mwbare$ to be determined in the following. The underlying idea is that the potential is generated by photon and $\omega$ states, where the $\omega$ is to be approximated by a narrow resonance, with parameters that correspond to renormalized versions of the bare parameters from Eq.~\eqref{ansatz}.

In order to determine these parameters, we first consider the $\ell^+\ell^-$ component
\beq
t_R(s)_{22}=-\frac{4\pi\alpha}{s}\big(1-\Pi_\ell(s)-\Pi_{\omega}(s)\big)^{-1},
\eeq
where we have used the fact that VP is the only correction that should appear in $\ell^+\ell^-$ scattering. In a narrow-width approximation for the $\omega$ one would expect 
\beq
\label{Pi_omega}
\Pi_{\omega}(s)=\frac{e^2\gwg^2s}{s-\mw^2+i\mw\Gw},
\eeq
where the $\omega$--$\gamma$ coupling has been chosen in agreement with the Lagrangian definition
\beq
\label{Lagr_omega_gamma}
\Lagr_{\omega\gamma}=-\frac{e}{2}\gwg F^{\mu\nu}\omega_{\mu\nu},\qquad \omega_{\mu\nu}=\partial_\mu\omega_\nu-\partial_\nu\omega_\mu.
\eeq
The potential in Eq.~\eqref{ansatz} produces
\begin{align}
\label{Pi_omega_potential}
\Pi_{\omega}(s)
&=-e^2 s\big(s-\mwbare^2\big)\frac{g_3^2}{\gw^2}\notag\\
&+\frac{e^2 s\big(\frac{g_3}{\gw}\big(s-\mwbare^2\big)\big)^2}{s-\mwbare^2+\Sw(s)\gw^2},
\end{align}
demonstrating that $g_3$ in Eq.~\eqref{ansatz} indeed needs to be accompanied by a factor $s$ to ensure gauge invariance.
The comparison to Eq.~\eqref{Pi_omega} then suggests the identification\footnote{The sign in $\gwg$ is motivated by the $2\pi$ channel and the definition of the $\rho$--$\omega$ mixing parameter, to match the sign conventions of Ref.~\cite{Hanhart:2016pcd}.}
\begin{align}
\label{couplings}
\Sw(s)&\equiv\Sw=\frac{\mwbare^2-\mw^2+i\mw\Gw}{\gw^2},\notag\\ 
\gwg&=-\frac{g_3}{\gw}\big(s-\mwbare^2\big).
\end{align}
The need to absorb an $s$ dependence into $\gwg$ shows that these relations are only strictly meaningful at the resonance mass, while polynomial ambiguities arise elsewhere. 
For the moment, we indicate this caveat by writing
\beq
\label{Pi_omega_poly}
\Pi_{\omega}(s)=P_\omega(s)+\frac{e^2\gwg^2s}{s-\mw^2+i\mw\Gw},
\eeq
with the role of the polynomial to be clarified in the following.

In total, the amplitudes in this two-channel system are
\begin{align}
\label{amplitudes_2channel}
 t_R(s)_{22}&=-\frac{e^2}{s}\big(1-\Pi(s)\big)^{-1},\\
 t_R(s)_{23}&=\frac{e^2\gw\gwg}{s-\mw^2+i\mw\Gw}\big(1-\Pi(s)\big)^{-1},\notag\\
 t_R(s)_{33}&=\bigg[\frac{\Pi_\omega(s)}{\Sw}-\frac{\gw^2(1-\Pi_\ell(s))}{s-\mw^2+i\mw\Gw}\bigg]\big(1-\Pi(s)\big)^{-1},\notag
\end{align}
where $\Pi(s)=\Pi_\ell(s)+\Pi_\omega(s)$.

\subsection{Comparison to $2\pi$ and $\ell^+\ell^-$}

The $2\pi$ channel could be treated in a similar way as $3\pi$ if the $\rho$ were assumed to be narrow. In this case, one would write in the vicinity of the pole~\cite{Hoferichter:2017ftn}  
\beq
F_\pi^V(s)=\frac{\gr\grg s_\rho}{s_\rho-s},
\eeq
with $\grg$ defined in analogy to $\gwg$ and $\gr$ related to the decay width via
\beq
\Gr=\frac{\gr^2\mr}{48\pi}\sigma_\pi^3(\mr^2). 
\eeq
If we then assume the integral in Eq.~\eqref{Pi_pi} to be dominated by contributions close to the resonance and expand in $\Gr$, we obtain
\begin{align}
\label{2pi_int_appr}
\Pi_\pi(s)&=-\frac{e^2\gr^2\grg^2}{48\pi}\frac{\mr}{\Gr}\sigma_\pi^3(\mr^2)\notag\\
&\quad\times\frac{s}{\pi}\int_{4\mpi^2}^\infty\diff s'\frac{\frac{\mr\Gr}{(s'-\mr^2)^2+(\mr\Gr)^2}}{s'-s-i\eps}.
\end{align}
The imaginary part of the integral can be read off from the $i\eps$ prescription, while in the narrow-width limit the numerator collapses to $\pi\delta(s'-\mr^2)$, producing
\beq
\label{Pi_pi_VMD}
\Pi_\pi(s)=\frac{e^2\grg^2 s}{s-\mr^2+i\mr\Gr},
\eeq
which coincides with the narrow-width expectation for $\Pi_\omega(s)$ in Eq.~\eqref{Pi_omega}. This indicates that ultimately the polynomial in Eq.~\eqref{Pi_omega_poly} should be put to zero.
In the same narrow-width approximation, the $e^+e^-\to\pi^+\pi^-$ peak cross section becomes
\begin{align}
\label{rho_cross_section}
\sigma(e^+e^-\to\pi^+\pi^-)\big(\mr^2\big)&=\frac{16\pi^2|\alpha(\mr^2)|^2\grg^2}{\mr\sigma_e(\mr^2)}\\
&\quad\times\frac{\Gr(\mr^2+2m_e^2)}{|s-\mr^2+i\mr\Gr|^2},\notag
\end{align}
which can be used to interpret the result for the $e^+e^-\to 3\pi$ cross section~\eqref{3pi_omega_cs} below.

\subsection{$e^+e^-\to 3\pi$ cross section}

To interpret results for the $3\pi$ channel we need to consider the cross section $e^+e^-\to 3\pi$. Its general form can be written as
\begin{align}
\label{cross_section}
  \sigma(e^+ e^- \to 3\pi)(s) &= \int_{s_-}^{s_+} \diff s' \int_{t_-}^{t_+} \diff t' \,\frac{\diff^2\sigma(s',t';s)}{\diff s' \, \diff t'},\notag\\
  \frac{\diff^2\sigma(s',t';s)}{\diff s' \, \diff t'} &= 
  \frac{\alpha^2(s't'u'-\mpi^2(s-\mpi^2)^2)}{192\pi s^3}\notag\\
  &\quad\times \big|\F(s',t',u';s) \big|^2\frac{s+2\me^2}{s\sigma_e(s)},
\end{align}
with $s'+t'+u'=s+3\mpi^2$ and integration boundaries
\begin{align}
 s_-&=4\mpi^2,\qquad s_+=(\sqrt{s}-\mpi)^2,\notag\\
 t_\pm&=\frac{1}{2}\big[s+3\mpi^2-s'\pm\sigma_\pi(s')\lambda^{1/2}(s,\mpi^2,s')\big].
\end{align}
The scalar function $\F(s',t',u';s)$ is normalized as
\beq
\label{Fnorm}
\F(0,0,0;0)=F_{3\pi}=\frac{1}{4\pi^2\Fpi^3},
\eeq
and decomposes as 
\begin{align}
\F(s',t',u';s)&=a(s)f_{3\pi}(s',t',u'),\notag\\
f_{3\pi}(s',t',u')&=\Omega_1^1(s')+\Omega_1^1(t')+\Omega_1^1(u'),
\label{deff3pi}
\end{align}
as long as rescattering effects are neglected~\cite{Niecknig:2012sj}. The simplest form of the normalization function consistent with Eq.~\eqref{Fnorm} and including the $\omega$ resonance reads
\beq
\label{defa}
a(s)=\frac{F_{3\pi}}{3}\bigg(1- c_{3\pi}\frac{s}{s-\mw^2+i\mw\Gw}\bigg),
\eeq
where VMD predicts $c_{3\pi}=1$. A similar form holds for the $3\pi$ decay width:
\begin{align}
 \Gw&=\frac{1}{128(2\pi)^3\mw^3}\int_{s_-}^{s_+}\diff s' \int_{t_-}^{t_+}\diff t'\\
 &\quad\times\Big(s't'u'-\mpi^2(\mw^2-\mpi^2)^2\Big)|c_\omega|^2\big|f_{3\pi}(s',t',u')\big|^2,\notag
\end{align}
where we have written the normalization in terms of another constant $c_\omega$, but assumed the same dependence on the Mandelstam variables as in Eq.~\eqref{deff3pi}. Numerically, this gives
\beq
\Gw\approx 2.865\times 10^{-7}\GeV^7 |c_\omega|^2.
\eeq
The peak cross section can then be expressed as 
\begin{align}
\sigma(e^+e^-\to3\pi)\big(\mw^2\big)&=\frac{16\pi^2\alpha^2F_{3\pi}^2|c_{3\pi}|^2}{9|c_\omega|^2\mw\sigma_e(\mw^2)}\notag\\
&\quad\times\frac{\Gw(\mw^2+2m_e^2)}{|s-\mw^2+i\mw\Gw|^2}.
\label{3pi_omega_cs} 
\end{align}
Matching to the expected form~\eqref{rho_cross_section} gives
\beq
|c_{3\pi}|=\frac{3|c_{\omega}|\gwg}{F_{3\pi}}\approx 0.98,
\eeq
demonstrating that, phenomenologically, $c_{3\pi}$ indeed comes out close to the VMD expectation. Finally, we remark that the form of the cross section~\eqref{3pi_omega_cs}  indeed matches onto $|t_R(s)_{23}|^2$ from Eq.~\eqref{amplitudes_2channel} as long as $\Gw\propto |\gw|^2$, which clarifies the role of this parameter. To work out the explicit relation one would need to assign $\xi_i$ and $\Gamma_i$ for $i=3\pi$, but since we will not consider $3\pi$ final states in the following, we do not pursue this avenue any further.

\subsection{Pseudoscalar decays}

In the formalism developed so far, the pseudoscalar decays of $P=\pi^0,\eta,\eta'$ into $\pi^+\pi^-\gamma$, $\ell^+\ell^-\gamma$, or $3\pi\gamma$ can be derived from Eq.~\eqref{def_Fi}, with a suitable choice of source terms $M_k$. To avoid the complications from $\rho$--$\omega$ mixing, we first consider the two-channel case $\{\ell^+\ell^-,3\pi\}$, before generalizing to the full system in~\ref{app:full}.

For the $3\pi$ channel, the ansatz~\eqref{MK_ansatz} applies, i.e., we can write
\beq
\label{M2}
M_3=a+\frac{\gw}{s-\mwbare^2}b+g_3 e^2 c,
\eeq
with constants $a$, $b$, $c$ to be determined in the following. Accordingly, to ensure consistency with Eq.~\eqref{ansatz}, we have
\beq
\label{M1}
M_2=c\frac{e^2}{s}.
\eeq
Using Eq.~\eqref{tR_def}, we find
\beq
F_i(s)=\big(\Gamma_\text{out}(s)\big)_i\big(\unity+t_R(s)\Sigma(s)\big)_{ij}M_j(s),
\eeq
in particular, 
\beq
F_2(s)=\frac{1}{1-\Pi(s)}\bigg(\frac{e^2}{s}c+\frac{\Pi_\omega(s)}{s g_3}\Big(a+\frac{\gw}{s-\mwbare^2}b\Big)\bigg).
\eeq
To interpret this form factor in the context of $P\to\ell^+\ell^-\gamma$ decays, we first consider the analog of Eq.~\eqref{pi+pi-_check} for a coupled-channel system of $\{\ell^+\ell^-,P\gamma\}$
\begin{align}
\label{pigamma_check}
\sigma(\ell^+\ell^-\to P\gamma)&=\frac{2\pi}{64\pi^2 s}\frac{1}{4}\frac{\sigma_{P\gamma}(s)}{\sigma_\ell(s)} \big|t(s)_{24}\big|^2\times\frac{2}{3}\\
&\equiv\frac{\pi^2\alpha^3}{6s}\frac{\sigma_{P\gamma}^3(s)\big|F_{P\gamma^*\gamma^*}(s,0)\big|^2}{|1-\Pi(s)|^2}\frac{\xi_\ell^2(s)}{\sigma_\ell(s)},\notag
\end{align}
where we defined the phase-space factor $\sigma_{P\gamma}(s)=(s-\MP^2)/s$. In addition, $V_R(s)$ is the same as in Eq.~\eqref{t_R_def} and $\Gamma_\text{out}(s)_{P\gamma}=e F_{P\gamma^*\gamma^*}(s,0)$, leaving only $\xi_{P\gamma}(s)$ to be determined. The matching in Eq.~\eqref{pigamma_check} yields
\beq
\xi_{P\gamma}^2(s)=\frac{1}{2}\big(s-\MP^2\big)^2.
\eeq
The same relation can also be derived from the $P\gamma$ contribution to the VP function $\Pi(s)$, supporting this assignment for a $P\gamma$ state, in terms of which the differential decay width becomes
\begin{align}
\label{Peeg}
\frac{\diff\Gamma(P\to \ell^+\ell^-\gamma)}{\diff s}&=\frac{\alpha^3(\MP^2-s)\xi_\ell^2(s)\xi_{P\gamma}^2(s)\sigma_\ell(s)}{12\MP^3s^2|1-\Pi(s)|^2}\notag\\
&\quad\times\big|F_{P\gamma^*\gamma^*}(s,0)\big|^2.
\end{align}
Due to the final-state photon also the three-particle phase space can be simplified, producing 
\begin{align}
\label{Peeg_ansatz}
 \frac{\diff\Gamma(P\to \ell^+\ell^-\gamma)}{\diff s}&=
 \frac{1}{(2\pi)^332\MP^3}\big(\MP^2-s\big)\sigma_\ell(s)\frac{1}{2}\notag\\
 &\quad\times\big|\M_{P\to \ell^+\ell^-\gamma}(s)\big|^2\times\frac{2}{3},
\end{align}
where we have written the general relativistic amplitude as $\M_{P\to \ell^+\ell^-\gamma}(s)$ in analogy to the scattering reactions studied before. The matching of Eqs.~\eqref{Peeg} and~\eqref{Peeg_ansatz} then determines
\begin{align}
\M_{P\to \ell^+\ell^-\gamma}(s)&=\frac{e^2}{s(1-\Pi(s))}\xi_\ell(s)\xi_{P\gamma}(s)e F_{P\gamma^*\gamma^*}(s,0)\notag\\
&\equiv \xi_\ell(s)\xi_{P\gamma}(s)eF_2(s),
\end{align}
where the last step gives the identification with the outcome of our coupled-channel formalism. The parameter $c=F_{P\gamma\gamma}$ thus determines the TFF normalization, and defining, in analogy to Eq.~\eqref{couplings}, $\tilde a=-\big(s-\mwbare^2\big)a/\gw$, we find the relation
\beq
F_{P\gamma^*\gamma^*}(s,0)=F_{P\gamma\gamma}+\Pi_\omega(s)\frac{\tilde a-b}{\gwg e^2}.
\eeq
This again strongly suggests to set $P_\omega(s)=0$ in Eq.~\eqref{Pi_omega_poly}, because then the matching
\beq
\tilde a-b\equiv -\frac{c_{3\pi}}{\gwg}F_{P\gamma\gamma}
\eeq
reproduces precisely the expected momentum dependence~\eqref{defa}. Our final result for the $P\to \ell^+\ell^-\gamma$ amplitude, related to the decay width by Eq.~\eqref{Peeg_ansatz}, reads
\begin{align}
\label{Meeg_2channel}
 \M_{P\to \ell^+\ell^-\gamma}(s)&=\frac{\xi_\ell(s)\xi_{P\gamma}(s)}{1-\Pi(s)}\frac{e^3 F_{P\gamma\gamma}}{s}\\
 &\quad\times \bigg(1-c_{3\pi}\frac{s}{s-\mw^2+i\mw\Gw}\bigg).\notag
\end{align}

\subsection{Radiative corrections to the $\omega$ pole parameters}

As final step before generalizing to the full system we consider the modifications to the $\omega$ parameters when including VP corrections. Such modifications arise because of
\begin{align}
1-\Pi_\ell(s)-\Pi_\omega(s)
&=\bigg[s-\mw^2+i\mw\Gw-\frac{e^2\gwg^2 s}{1-\Pi_\ell(s)}\bigg]\notag\\
&\quad\times\frac{1-\Pi_\ell(s)}{s-\mw^2+i\mw\Gw}.
\end{align}
Expanding the first factor in $e$,
\begin{align}
&s-\mw^2+i\mw\Gw-\frac{e^2\gwg^2 s}{1-\Pi_\ell(s)}\notag\\
&=s\big(1-e^2\gwg^2\big)-\mw^2+i\mw\Gw+\Order\big(e^4\big)\notag\\
&\equiv\big(1-e^2\gwg^2\big)\bigg(s-\mwb^2+i\mwb\Gwb\bigg)+\Order\big(e^4\big),
\end{align}
we find that the new $\omega$ parameters are given by
\begin{align}
\mwb&=\bigg(1+\frac{e^2\gwg^2}{2}\bigg)\mw+\Order(e^4),\notag\\
\Gwb&=\bigg(1+\frac{e^2\gwg^2}{2}\bigg)\Gw+\Order(e^4),
\end{align}
and the couplings are renormalized according to
\beq
\gwb=\gw\sqrt{Z},\qquad \gwgb=\gwg\sqrt{Z},\qquad Z=1+e^2\gwg^2.
\eeq
Numerically, these shifts are rather small, with 
$Z-1=3.3\times 10^{-4}$, i.e., the change in $\gwg$ is completely negligible. As concerns the pole parameters~\cite{Hoferichter:2019mqg,Hoid:2020xjs},
\begin{align}
\label{omega_parameters_radiative_corrections}
\Delta\mw&=\mwb-\mw=0.13\MeV,\notag\\
\Delta\Gw&=\Gwb-\Gw=1.4\keV,
\end{align}
the shift in the mass is comparable to current experimental uncertainties, while the main effect in the width comes instead from the interaction with the $2\pi$ channel. Using the approximation~\eqref{Pi_pi_VMD}, one finds
\begin{align}
\label{omega_width}
\Delta\Gw&=\frac{e^2\gwg^2}{2}\Gw+\frac{\mw^2}{\Gr-\Gw}e^2\grg^2\big(e^2\gwg^2-2\epsrw\big)\notag\\
&=-0.06\MeV.
\end{align}

\section{Full system}
\label{app:full}

In this section, we present our final results for the full system $\{2\pi,\ell^+\ell^-,3\pi\}$ with $\gamma$ and $\omega$ mediators in the resonance potential. We write the expressions in terms of the VP-subtracted $\omega$ parameters $\mw$, $\Gw$, with the understanding that they are related to the physical pole parameters including VP by means of Eqs.~\eqref{omega_parameters_radiative_corrections} and~\eqref{omega_width}. This convention simplifies expressions, and at the same time makes the factorization of VP manifest. We show the results for a single lepton species $\ell$, with straightforward generalization to multiple generations. 

\subsection{Scattering channels}

The ansatz for the full potential reads
\begin{align}
\label{ansatz_full}
V_R(s)&=-\frac{4\pi\alpha}{s}\begin{pmatrix} 1 & 1 & g_3 s\\ 1 & 1 & g_3 s\\g_3 s &  g_3 s & (g_3 s)^2 \\ \end{pmatrix} 
\notag\\
&- \frac{1}{s-\mwbare^2}\begin{pmatrix} \gww^2  & 0 & \gww\gw\\ 0 &  0 & 0 \\\gww\gw & 0 & \gw^2 \\ \end{pmatrix},\notag\\
\Sigma(s)&=\text{diag}\big(\Sigma_\pi(s),\Sigma_\ell(s),\Sw(s)\big),
\end{align}
where $\gww$ parameterizes the isospin-breaking coupling of the $2\pi$ channel to the $\omega$. In practice, we will neglect higher orders in $\gww$, given that at this level also other subleading effects, such as the $\omega\to\pi^0\gamma$ channel, would need to be considered. The full VP function becomes
\begin{align}
\Pi(s)&=\Pi_\ell(s)+\Pi_\pi(s)\bigg(1+\frac{2s\eps_{\rho\omega}}{\mw^2-s-i\mw\Gw}\bigg)\notag\\
&+\Pi_\omega(s)+\Order\big(\gww^2\big),
\end{align}
with $\Pi_\omega(s)$ as defined in Eq.~\eqref{Pi_omega}, and the $\rho$--$\omega$ mixing parameter
\beq
\epsrw=\gww\gwg.
\eeq
The full amplitudes are
\begin{align}
\label{hattR}
 t_R(s)&=\frac{1}{1-\Pi(s)}\hat t_R(s)+\Order\big(\gww^2\big),\notag\\
 \hat t_R(s)_{11}&=-\frac{e^2}{s}\bigg(1+\frac{2\epsrw s}{\mw^2-s-i\mw\Gw}\bigg) \notag\\
 &+ \frac{\epsrw^2}{\gwg^2}\frac{1}{\mw^2-s-i\mw\Gw},\notag\\
 \hat t_R(s)_{12}&=-\frac{e^2}{s}\bigg(1+\frac{\epsrw s}{\mw^2-s-i\mw\Gw}\bigg),\notag\\
 \hat t_R(s)_{22}&=-\frac{e^2}{s},\notag\\
 \hat t_R(s)_{13}&=\frac{e^2\gw\gwg}{s-\mw^2+i\mw\Gw}\bigg(1-\frac{\epsrw(1-\Pi_\ell(s))}{e^2\gwg^2}\bigg),\notag\\
 \hat t_R(s)_{23}&=\frac{e^2\gw\gwg}{s-\mw^2+i\mw\Gw}\bigg(1-\frac{\epsrw \Pi_\pi(s)}{e^2\gwg^2}\bigg),\notag\\
 \hat t_R(s)_{33}&=\frac{\Pi_\omega(s)}{\Sw}\bigg(1-\frac{2\epsrw \Pi_\pi(s)}{e^2\gwg^2}\bigg)\notag\\
 &-\frac{\gw^2(1-\Pi_\ell(s)-\Pi_\pi(s))}{s-\mw^2+i\mw\Gw}.
\end{align}
$t_R(s)_{23}$ and $t_R(s)_{33}$ indeed generalize Eq.~\eqref{amplitudes_2channel}, and in the conventions~\eqref{couplings} the sign in the $\rho$--$\omega$-mixing correction in $e^+e^-\to\pi^+\pi^-$ agrees with Ref.~\cite{Hanhart:2016pcd}.
Note that the expansion in $\gww$ is clearly not appropriate for the isospin-violating corrections to $\pi^+\pi^-$ scattering, where terms of order $\Order(\gww^2)$ should be counted at the same order as $\Order(e^2 \gww)$. For completeness, we have therefore retained the corresponding correction in $t_R(s)_{11}$.

\subsection{Decays and transition form factors}
\label{app:decays+TFFs}

Generalizing Eqs.~\eqref{M2} and~\eqref{M1}, we make the following ansatz for the source terms
\beq
M(s)=c\frac{e^2}{s}\begin{pmatrix}
                    1\\ 1\\ g_3 s\\
                   \end{pmatrix}
                   +\frac{b}{s-\mwbare^2}
                   \begin{pmatrix}
                                          \gww\\0\\ \gw\\
                                         
                                         \end{pmatrix}
                                         +\begin{pmatrix}
                                           -P(s)\\0\\a\\
                                          \end{pmatrix},
\eeq
with parameters $a$, $b$, $c$ to be interpreted as before, and only the role of the function $P(s)$ to be clarified. As a first step, the result for the $P\to\ell^+\ell^-\gamma$ amplitude generalizes to
\begin{align}
 \M_{P\to \ell^+\ell^-\gamma}(s)&=\frac{\xi_\ell(s)\xi_{P\gamma}(s)}{1-\Pi(s)}\frac{e^3}{s}
 F_{P\gamma^*\gamma^*}(s,0),\notag\\
 F_{P\gamma^*\gamma^*}(s,0)&=F_{P\gamma\gamma}+P(s)\Sigma_\pi(s)\notag\\
 &\quad\times\bigg(1+\frac{\epsrw s}{\mw^2-s-i\mw\Gw}\bigg)\notag\\
 &+\frac{F_{P\gamma\gamma}w_{P\omega\gamma} s}{\mw^2-s-i\mw\Gw}\bigg(1+\frac{\epsrw}{s \gwg^2}\Sigma_\pi(s)\bigg)\notag\\
 &+\Order\big(e^2\epsrw\big).
 \label{MPllg}
\end{align}
Here, we have eliminated the parameter $c_{3\pi}$ in favor of the $\omega$ contribution to the slope of the TFF\footnote{For simplicity, we only consider $P=\eta,\eta'$, otherwise, the different isospin structure for the $\pi^0$ TFF would need to be taken into account. In particular, Eq.~\eqref{weights} produces $w_{\pi^0\omega\gamma}\approx 0.5$, consistent with $c_{3\pi}=2w_{\pi^0\omega\gamma}$ in this case.}
\beq
c_{3\pi}\to \mw^2 b_P^{\omega}=\frac{\mw^2}{F_{P\gamma\gamma}}\frac{\diff F_{P\gamma^*\gamma^*}^{\omega}(s,0)}{\diff s}\bigg|_{s=0}=w_{P \omega\gamma},
\eeq
where the weights are given by~\cite{Gan:2020aco}
\begin{align}
\label{weights}
 w_{PV\gamma}^2=
 \begin{cases}
 \frac{9M_V^2\MP^3\Gamma(V\to e^+e^-)\Gamma(V\to P\gamma)}{2\alpha(M_V^2-\MP^2)^3\Gamma(P\to\gamma\gamma)}\,\,\,\, \text{if}\,M_V>\MP,\\
 \frac{3\MP^6\Gamma(V\to e^+e^-)\Gamma(P\to V\gamma)}{2\alpha M_V(\MP^2-M_V^2)^3\Gamma(P\to\gamma\gamma)}\quad \,\,\,\,\,\text{if}\,\MP>M_V.
 \end{cases}
\end{align}
Determining the signs by the comparison to VMD~\cite{Gan:2020aco}, the numerical values based on Table~\ref{tab:fit_input} are 
\begin{align}
\label{eq:isoscalar_coup}
w_{\eta'\omega\gamma}&=0.072(2),&
w_{\eta'\phi\gamma}&=0.158(6),
\end{align}
and in terms of couplings $g_{PV\gamma}$,
\begin{align} \label{gPVgamma}
g_{PV\gamma}^2=\begin{cases}
\frac{24\Gamma(V\to P\gamma)}{\alpha}\Big(\frac{M_V}{M_V^2-\MP^2}\Big)^3\quad \text{if}\,M_V>\MP,\\
                \frac{8\Gamma(P\to V\gamma)}{\alpha}\Big(\frac{\MP}{\MP^2-M_V^2}\Big)^3\quad \,\text{if}\,\MP>M_V,
               \end{cases}
\end{align}
one finds 
\beq
\label{w_g}
w_{PV\gamma}^2=\frac{g_{PV\gamma}^2g_{V\gamma}^2}{F_{P\gamma\gamma}^2}.
\eeq
In analogy to Eq.~\eqref{Peeg_ansatz}, we have
\begin{align}
 \frac{\diff\Gamma(P\to \pi^+\pi^-\gamma)}{\diff s}&=
 \frac{1}{(2\pi)^332\MP^3}\big(\MP^2-s\big)\sigma_\pi(s)\frac{1}{2}\notag\\
 &\quad\times  \big|\M_{P\to \pi^+\pi^-\gamma}(s)\big|^2\times \frac{2}{3},
\end{align}
with 
\begin{align}
\M_{P\to \pi^+\pi^-\gamma}(s)&=-eF_\pi^V(s)\xi_\pi(s)\xi_{P\gamma}(s)\bigg( - \frac{e^2 F_{P\gamma\gamma}}{s}\notag\\
&+\frac{F_{P\gamma\gamma}w_{P\omega\gamma}}{\gwg^2}\frac{\epsrw-e^2\gwg^2}{\mw^2-s-i\mw\Gw}\notag\\
&+P(s)\big(1+\Pi_\pi(s)\big)+\Order\big(e^4\big)\bigg),
\end{align}
where the coefficient of the $\omega$ admixture can be written as $-g_{P\omega\gamma}\epsrw/\gwg$ due to Eqs.~\eqref{eq:couplings_gog_geog}~and~\eqref{w_g}. 
The resulting spectrum becomes
\begin{align}
\label{spectrum_pipigamma_app}
 \frac{\diff\Gamma(P\to \pi^+\pi^-\gamma)}{\diff s}&=16\pi\alpha\Gamma_0|F_\pi^V(s)|^2 \bigg|P(s)\big(1+\Pi_\pi(s)\big)\notag\\
 - \frac{e^2 F_{P\gamma\gamma}}{s}&-\frac{g_{P\omega\gamma}}{\gwg}\frac{\epsrw-e^2\gwg^2}{\mw^2-s-i\mw\Gw}\bigg|^2,\notag\\
 \Gamma_0&=\frac{2s}{3}\bigg(\frac{\MP^2-s}{16\pi\MP}\sigma_\pi(s)\bigg)^3,
\end{align}
as given in Eq.~\eqref{spectrum_pipigamma},
in agreement with Refs.~\cite{Kubis:2015sga,Hanhart:2016pcd} upon the identification $2P(s)=A(1+\beta s+\gamma s^2)$ (and separating the factor $e^2=4\pi\alpha$) and taking into account the sign of $g_{\eta' \omega \gamma}$ in Eq.~\eqref{eq:couplings_gog_geog}. The function $P(s)$ thus emerges as a parameterization of the $P\to\pi^+\pi^-\gamma$ spectrum. Higher orders in $e^2$ have been included primarily to ensure that the definition of $\epsrw$ agrees exactly with the conventions in $e^+e^-\to\pi^+\pi^-$~\cite{Colangelo:2022prz}; the shift $\epsrw\to \epsrw-e^2\gwg^2$ in the numerator of the $\omega$ propagator corresponds to the photon contribution in $\epsrw$ as defined in resonance chiral perturbation theory~\cite{Urech:1995ry,Bijnens:1996nq,Bijnens:1997ni}. 

\subsection{Dispersion relations}

The previous discussion now allows us to write down dispersion relations for the pseudoscalar TFF including a consistent treatment of $\rho$--$\omega$ mixing. Reading off the discontinuities from Eq.~\eqref{MPllg},
\begin{align}
\label{discontinuity}
 \Impipi F_{P\gamma^*\gamma^*}(s,0)&=\frac{s}{48\pi}\sigma_\pi^3(s)|F_\pi^V(s)|^2\bigg[P(s)\notag\\
 &\quad\times\bigg(1+\frac{s \epsrw}{\mw^2-s-i\mw\Gw}\bigg)^{*}\notag\\
 &+\frac{F_{P\gamma\gamma}w_{P\omega\gamma} \epsrw}{\gwg^2}\frac{1}{\mw^2-s-i\mw\Gw}\bigg],\notag\\
 \Impipipi F_{P\gamma^*\gamma^*}(s,0)&=\pi s\delta(s-\mw^2)\bigg[F_{P\gamma\gamma}w_{P\omega\gamma}\notag\\
 &\quad\times\bigg(1+\frac{\epsrw}{\gwg^2 s}\Sigma_\pi(s)\bigg)^{*}\notag\\
 &+\epsrw P(s)\Sigma_\pi(s)\bigg],
\end{align}
we first observe that, as expected, the sum of the double discontinuities vanishes:
\begin{align}
\label{double_cancel}
\Im\big[\Impipi F_{P\gamma^*\gamma^*}(s,0)\big]&=-\Im\big[\Impipipi F_{P\gamma^*\gamma^*}(s,0)\big]\notag\\
&=\frac{s \epsrw}{48}\sigma_\pi^3(s)|F_\pi^V(s)|^2\delta(s-\mw^2)\notag\\
&\quad\times \bigg(\frac{F_{P\gamma\gamma}w_{P\omega\gamma}}{\gwg^2}-s P(s)\bigg).
\end{align}
Next, we replace
\beq
P(s)\Sigma_\pi(s)\to\frac{s}{48\pi^2}\int_{4\mpi^2}^\infty\diff s'\frac{\sigma_\pi^3(s')P(s')|F_\pi^V(s')|^2}{s'-s-i\eps},
\eeq
which does not affect the cancellation of the double discontinuities, but ensures that all dispersive integrals related to $\Sigma_\pi(s)$ have the same convergence properties. In the end, this step amounts to a choice of subtraction scheme, which, in the coupled-channel formalism, corresponds to polynomial ambiguities. 

To derive a dispersive representation for the TFF, we start from the once-subtracted version
\begin{align}
F_{P\gamma^*\gamma^*}(s,0)&=F_{P\gamma\gamma}+\frac{s}{\pi}\int_{4\mpi^2}^\infty\diff s'\notag\\
&\quad\times\frac{\sum_{i=\pi\pi,3\pi}\text{Im}_{i} F_{P\gamma^*\gamma^*}(s',0)}{s'(s'-s)}.
\end{align}
To keep track of the $\omega$ propagator consistently at all stages of the calculation, it is useful to replace
\beq
\frac{s}{\mw^2-s-i\mw\Gw}\to\frac{s}{\pi}\int\diff s'\frac{\pi \delta(s'-\mw^2)}{s'-s-i\eps}
\eeq
and only restore the finite width in the end.\footnote{This can be done more rigorously by using a smeared-out version of the $\delta$-function instead, but the results are identical. See also the discussion leading to Eq.~\eqref{Pi_pi_VMD}.}
Combining denominators in the various dispersion integrals, one can show that indeed the $\omega$ propagator factorizes, leading to a form very similar to Eq.~\eqref{MPllg}:
\begin{align}
\label{rep_final_app}
 F_{P\gamma^*\gamma^*}(s,0)&=F_{P\gamma\gamma}+
 \bigg[1+\frac{\epsrw s}{\mw^2-s-i\mw\Gw}\bigg]\notag\\
 &\quad\times \frac{s}{48\pi^2}\int_{4\mpi^2}^\infty\diff s'\frac{\sigma_\pi^3(s')P(s')|F_\pi^V(s')|^2}{s'-s-i\eps}\notag\\
 &+\frac{F_{P\gamma\gamma}w_{P\omega\gamma}s}{\mw^2-s-i\mw\Gw}\bigg[1+\frac{\epsrw s}{48\pi^2\gwg^2}
 \notag\\
 &\quad\times\int_{4\mpi^2}^\infty\diff s'\frac{\sigma_\pi^3(s')|F_\pi^V(s')|^2}{s'(s'-s-i\eps)}\bigg].
\end{align}
This is our final result for a dispersive representation of $F_{P\gamma^*\gamma^*}(s,0)$ that includes corrections from $\rho$--$\omega$ mixing in a consistent manner, with crucial ingredient the function $P(s)$ as determined by the $P\to \pi^+\pi^-\gamma$ spectrum~\eqref{spectrum_pipigamma}. In particular, from the limit $s\to\infty$ one can read off the sum rule
\begin{align}
\label{sum_rule}
  F_{P\gamma\gamma}&=\frac{1-\epsrw}{48\pi^2}\int_{4\mpi^2}^\infty\diff s'\,\sigma_\pi^3(s')P(s')|F_\pi^V(s')|^2\\
 &+F_{P\gamma\gamma}w_{P\omega\gamma}\notag\\
 &\quad\times\bigg[1-\frac{\epsrw}{48\pi^2\gwg^2}\int_{4\mpi^2}^\infty\diff s'\frac{\sigma_\pi^3(s')|F_\pi^V(s')|^2}{s'}\bigg].\notag
\end{align}
The isoscalar part of the TFF also involves a contribution from the $\phi$, which simply amounts to adding 
\beq
\label{eq:phi_contribution}
F_{P\gamma^*\gamma^*}^\phi(s,0)=\frac{F_{P\gamma\gamma}w_{P\phi\gamma}s}{\mphi^2-s-i\mphi\Gphi}
\eeq
to the right-hand side of Eq.~\eqref{rep_final_app} (and $F_{P\gamma\gamma}w_{P\phi\gamma}$ to the one of Eq.~\eqref{sum_rule}).
The combination of Eqs.~\eqref{rep_final_app} and \eqref{eq:phi_contribution} is the form \eqref{rep_final} given in the main text.

As a final cross check, we show that Eq.~\eqref{rep_final_app} reproduces the correct VMD limit. To this end, we first approximate the dispersive integrals by their narrow-width limit, in analogy to Eq.~\eqref{Pi_pi_VMD}. The second integral already takes VMD form, while for the former we need $|P(M_\rho^2)|$. This quantity can be related to $\Gamma(P\to\rho \gamma)$ by integrating Eq.~\eqref{spectrum_pipigamma} and again replacing the integral by its narrow-width limit. In this way, we can remove $|P(M_\rho^2)|$ in favor of $w_{P\rho\gamma}$, see Eq.~\eqref{weights}, and using the VMD relations $g_{\rho\pi\pi}=1/\grg$, $\grg=3\gwg$, we obtain 
\begin{align}
\label{rep_final_VMD}
 F_{P\gamma^*\gamma^*}^\text{VMD}(s,0)&=F_{P\gamma\gamma}\bigg[1+
 \bigg(1+\frac{\epsrw s}{\mw^2-s-i\mw\Gw}\bigg)\notag\\&\quad\times\frac{w_{P\rho\gamma}s}{\mr^2-s-i\mr\Gr}\notag\\
 &+\frac{w_{P\omega\gamma}s}{\mw^2-s-i\mw\Gw}\notag\\
 &\quad\times\bigg(1+\frac{9\epsrw s}{\mr^2-s-i\mr\Gr}\bigg)\bigg],
\end{align}
in agreement with the expected VMD result.

\bibliographystyle{utphysmod.bst}
\balance
\bibliography{refs}

\end{document}